# BIG DATA GENERATED BY CONNECTED AND AUTOMATED VEHICLES FOR SAFETY MONITORING, ASSESSMENT AND IMPROVEMENT

## FINAL REPORT, YEAR 3

SOUTHEASTERN TRANSPORTATION CENTER

Asad Khattak, Iman Mahdinia, Sevin Mohammadi, Amin Mohammadnazar & Behram Wali

April 2019

**US DEPARTMENT OF TRANSPORTATION GRANT DTRT13-G-UTC34**

1. Report No.
STC-2019-M4.UTK

2. Government Accession No.

3. Recipient's Catalog No.

4. Title and Subtitle
Big Data Generated by Connected and Automated Vehicles for Safety Monitoring, Assessment and Improvement

5. Report Date
April 2019

6. Source Organization Code
N/A

7. Author(s)
Asad Khattak, Iman Mahdinia, Sevin Mohammadi, Amin Mohammadnazar & Behram Wali

8. Source Organization Report No.
STC-2019-M4.UTK

9. Performing Organization Name and Address
Southeastern Transportation Center
309 Conference Center Building
Knoxville, Tennessee 37996-4133
865.974.5255

10. Work Unit No. (TRAIS)

11. Contract or Grant No.
DTRT12-G-UTC34

12. Sponsoring Agency Name and Address

US Department of Transportation
Office of the Secretary of Transportation Research
1200 New Jersey Avenue, SE Washington, DC 20590

13. Type of Report and Period Covered
Final Report:
August 2016-March 2019

14. Sponsoring Agency Code
USDOT/OST-R

15. Supplementary Notes: None

16. Abstract
This report focuses on safety aspects of connected and automated vehicle (CAV). The fundamental question to be answered is how can CAVs improve road users' safety? Using advanced data mining and thematic text analytics tools, the goal is to systematically synthesize studies related to Big Data for safety monitoring and improvement. Within this domain, the report systematically compares Big Data initiatives related to transportation initiatives nationally and internationally and provides insights regarding the evolution of Big Data science applications related to CAVs and new challenges. The objectives addressed are:

- Creating a database of Big Data efforts by acquiring reports, white papers, and journal publications;
- Applying text analytics tools to extract key concepts, and spot patterns and trends in Big Data initiatives;
- Understanding the evolution of CAV Big Data in the context of safety by quantifying granular taxonomies and modeling entity relations among contents in CAV Big Data research initiatives, and
- Developing a foundation for exploring new approaches to tracking and analyzing CAV Big Data and related innovations.

The study synthesizes and derives high-quality information from innovative research activities undertaken by various research entities through Big Data initiatives. The results can provide a conceptual foundation for developing new approaches for guiding and tracking the safety implications of Big Data and related innovations.

18. Distribution Statement

Unrestricted

| 19. Security Classif. (of this report) | 20. Security Classif. (of this page) | 21. No. of Pages | 22. Price |
|---|---|---|---|
| Unclassified | Unclassified | 53 | N/A |

## DISCLAIMER

## ACKNOWLEDGEMENT

Data used for this project are from several sources, including peer-reviewed journal papers, white papers, conference papers, reports, and theses collected through search engines such as Google Scholar, Science Direct, Web of Science, and journal databases. This project used software packages R and QDA Miner for data processing and visualization.

The research was supported through the Southeastern Transportation Center, sponsored by the United States Department of Transportation through grant number DTRT13-G-UTC34. Special thanks are extended to the following entities for their support: The Transportation Engineering & Science Program, the Initiative for Sustainable Mobility, and colleagues from the University of Kentucky and the University of Central Florida who also worked on the Southeastern Transportation Center's Big Data major research initiative.

The support from Dr. Richards and Ms. DeAnna Flinchum at the University of Tennessee is gratefully acknowledged. Dr. Melany Noltenius edited an earlier draft of this report.

A related paper was presented during the 2019 Transportation Research Board Annual Meeting.

Shay E., A. Khattak, A Boggs, Safety in the Connected and Automated Vehicle Era: A US Perspective on Research Needs, Transportation Research Board, 98th Annual Meeting, 2019.

We are very grateful to Dr. Elizabeth Shay and Ms. Ali Boggs for their contributions.

The views expressed in the report are those of the authors, who are responsible for the facts and accuracy of information presented herein.

## TABLE OF CONTENTS

## LIST OF FIGURES

## LIST OF TABLES

## Executive Summary

Rapid technological advancements in recent years have established the elemental foundation of cooperative vehicle control systems, manifesting in emerging connected and automated vehicles (CAV). The next frontier of transportation development is to equip motor vehicles and transportation systems with wireless communication technologies in a bid to establish cooperative, well informed, intelligent and proactive transportation systems. In this context, it is important to consider transformational advancements in transportation related data collection techniques that acquire increasing amounts of information generated by electronic sensors in combined ecosystems of CAVs. While generated large-scale empirical data has significant potential to facilitate deeper understanding of transportation-related problems, integrating and processing large datasets in a meaningful manner is still an open challenge. To this end, the Southeastern Transportation Center's (STC) "Big Data" major research initiative, which focuses on developing innovative frameworks for generating useful knowledge from fragmented, disorganized, and difficult-to-analyze "Big Data," is of critical importance.

The fundamental question to be answered is how CAVs will improve road users' safety? On the one hand, human errors that contribute to a majority of crashes can be potentially avoided with CAVs, resulting in safer transportation system operation. On the other hand, the safety of vehicle automation systems is still being researched and tested. Higher level automation is in early stages of development. Some drivers are uncomfortable with relinquishing control of their vehicles, especially in complex urban environments. At vehicular speeds, occupants' lives can depend on the performance of sensors and the big data that is being generated and transmitted. Hence, creating, processing, and analyzing big data generated by vehicles in real-time is critical for safe operation of CAVs.

This report focuses on extracting useful information from studies that relate to data generated by CAVs. By using advanced data mining and thematic text analytics tools, the goal is to systematically synthesize studies related to Big Data for safety monitoring and improvement. Within this domain, the report systematically compares Big Data initiatives nationally and internationally and provides insights regarding the evolution of Big Data science applications related to CAVs and new challenges. The key objectives are to:

- Create a repository of Big Data efforts by acquiring reports, white papers, and journal publications.
- Apply text analytics tools to extract key concepts, spot patterns and trends in Big Data initiatives.
- Understand the evolution of CAV Big Data in the context of safety by quantifying granular taxonomies and modeling entity relations among contents in other CAV Big Data research initiatives.
- Develop a foundation for exploring new approaches for tracking and analyzing CAV Big Data and related innovations.

The study synthesizes and derives high-quality information from innovative research activities undertaken by various research entities through Big Data initiatives. The results can provide a conceptual foundation for developing new approaches to guide and track safety implications of Big Data and related innovations.

# 1. Introduction

Large and varied data sets or "Big Data" enabled by digital connectivity between vehicles and transportation infrastructure are being used to help vehicles navigate in real time and they are being archived for further analysis. They allow vehicles, drivers, and traffic managers to communicate, coordinate and cooperate, potentially enhancing traffic safety through warnings, alerts, control assists, and at high levels of automation taking over control of the vehicle. However, the complex process of examining and extracting useful information from the multitude of data streams and datasets is in its infancy, and uncovering information involving hidden patterns and unknown correlations (increasingly in real-time) is difficult. Related to these issues, we outline the promise of connected and automated vehicles and issues associated with connected and automated vehicles and Big Data; and we describe the research objectives and approaches that are emerging in literature.

## 1.1. Problem Statement

Rapid technological advancements in recent years have established the elemental foundation of Cooperative Intelligent Transportation Systems (C-ITS), manifesting in emerging connected and automated vehicles (CAVs). The next frontier of transportation development is to equip motor vehicles and transportation systems with wireless communication technologies in a bid to establish cooperative, well informed, and proactive transportation systems. In this context, it is important to consider transformational advancements in transportation related data collection techniques that acquire increasing amounts of information generated by electronic sensors in combined ecosystems of CAVs. While generated large-scale empirical data has significant potential to facilitate deeper understanding of transportation-related problems, integrating and processing large datasets in a meaningful manner is still an open challenge. To this end, the Southeastern Transportation Center's (STC) "Big Data" major research initiative, which focuses on developing innovative frameworks for generating useful knowledge from fragmented, disorganized and difficult to analyze "Big Data" is of critical importance.

The fundamental question to be answered is how CAVs will improve road users' safety? In this regard, the study explores if the literature provides insights on the potential to avoid human errors when control is transferred to CAVs. The study also explores literature on how big data can be harnessed for safe operation of CAVs and relevant challenges.

During the first two years of the Big Data Major Research Initiative (MRI), STC member universities undertook and completed significant and innovative research activities, which collectively generated new knowledge for understanding and developing Big Data driven solutions to address safety problems. Such work has been successfully published in high-impact international journals, and continuing efforts are focused on expanding the domain of using Big Data for addressing, not only safety challenges, but also mobility and environmental concerns (1-9). The continuing progress achieved in Big Data analytics under this MRI is crucial, but there is also an exigency to track the STC's Big Data innovations, which can ultimately encourage a holistic understanding of the evolution of Big Data science and its implications for transportation issues. In addition, because of increases in national and international research activities related to Intelligent Transportation Systems and CAVs, it is crucial to systematically synthesize and extract valuable and high-quality knowledge from other Big Data initiatives and publications. New knowledge generated under the STC Big Data MRI together with other initiatives can ultimately help in developing a foundation for guiding future research initiatives and developing new procedures for tracking and analyzing CAV Big Data and related innovations.

### 1.2. Research Objectives

By using advanced data mining and thematic text analytics tools, the goal of this research is to systematically synthesize the work done by STC related to the utilization of Big Data generated by connected and automated vehicles for safety monitoring and improvement. Within this domain of Big Data, the project also systematically compares STC's Big Data initiatives with broader research work conducted under other major initiatives (nationally and internationally), as well as, explores the evolving field of Big Data related to CAV. The key objectives are to:

- Develop a comprehensive database of STC's CAV Big Data related efforts as well as other Big Data initiatives by acquiring reports, white papers, and journal publications;
- Apply rigorous and advanced machine learning text analytics tools that extract key concepts, spot patterns and trends in STC sponsored Big Data initiatives;
- Understand the evolution of CAV Big Data by quantifying granular taxonomies and modeling entity relations among contents in other national and international CAV Big Data research initiatives, and
- Develop a foundation for exploring new approaches to tracking and analyzing CAV Big Data and related innovations.

### 1.3. Research Approach

In the third year of this Big Data MRI, the overall theme of activities focusses on conducting rigorous and advanced text analyses of research activities undertaken in STC's Big Data initiatives and related national and international efforts. Research focuses on deriving high-quality structured information from safety oriented Big Data research. This helps investigators systematically structure and create a synopsis of the current body of Big Data knowledge used to extract meaningful information and produce new insights by spotting Big Data innovation indicators. Figure 1 presents the general conceptual framework for text mining, while Table 1 briefly explains various methods and key outcomes in the context of CAV Big Data.

Advanced text analytics facilitates systematic analysis of otherwise "random textual data" through the application of statistical pattern learning techniques (Figure 1). This study categorizes and stems a corpora of individual research articles, reports, and white papers (under STC and other initiatives) to spot unique concepts and entities to produce granular taxonomies (Figure 1). Specifically, text categorization and stemming techniques help extract the most important key topics and phrases through factor analysis (10). As shown in Figure 1, it is relevant to apply machine learning natural language processing techniques (in addition to statistical procedures) in order to uncover the hidden thematic structure in a collection of textual information, in this case individual papers or reports (11).

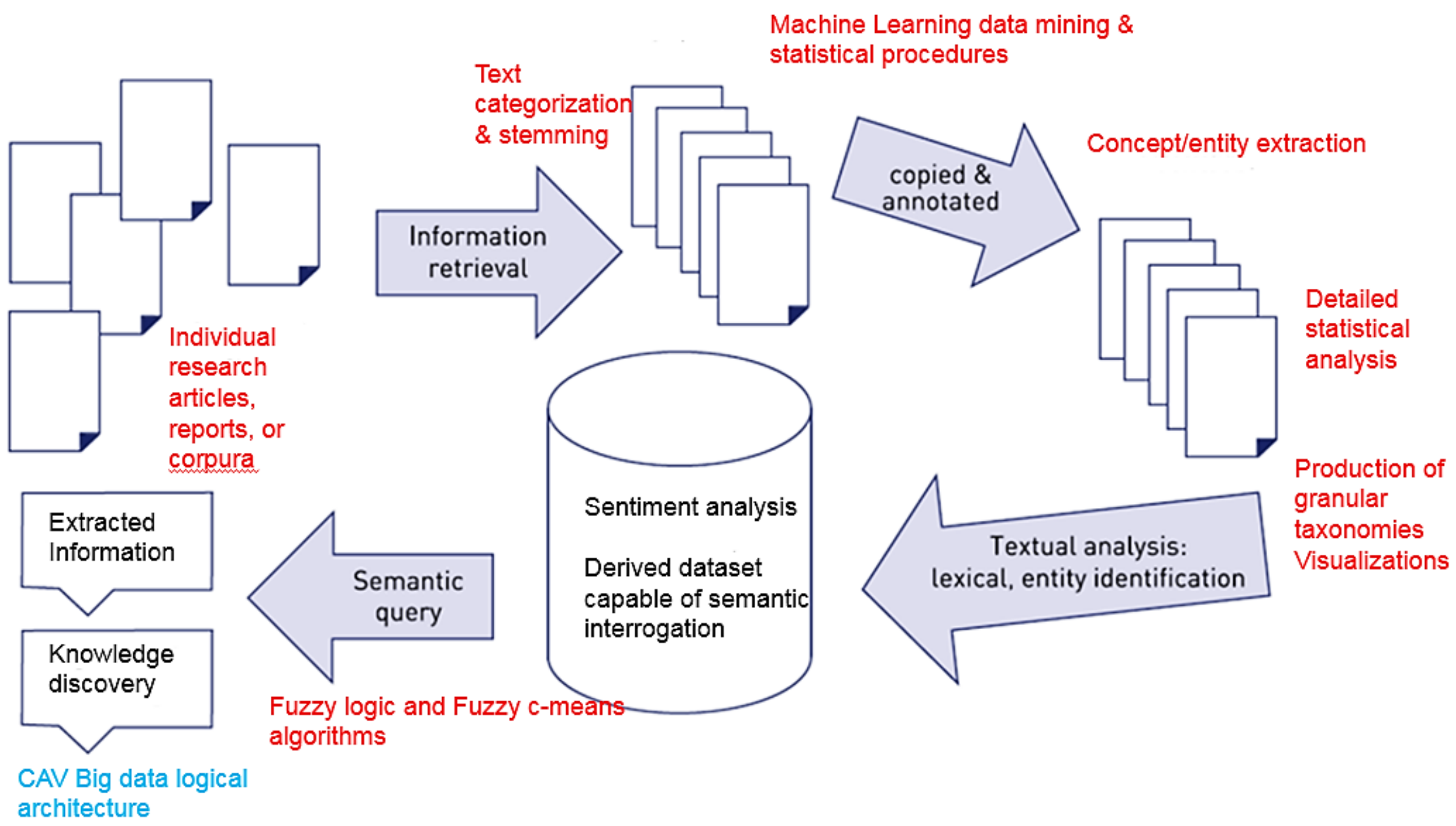

**Figure 1: Scheme of the Text Mining Framework**

After extracting key topics and phrases, researchers conducted detailed statistical analysis that spots case occurrences. Specifically, while controlling for similarity/correlation among key topics, researchers conducted hierarchical cluster analysis (1*2)* and multidimensional scaling to understand the current evolution of CAV Big Data science (13). As opposed to traditional measures of co-occurrence, this study applies probabilistic corrections to standard correlation indexes to account for the possibility that two topics can sometimes co-occur by chance.

Various visualization tools such as Dendrograms and 2D/3D concept maps of key concepts based on co-occurrences display the results obtained from statistical analysis. Note that hierarchical cluster analysis and multidimensional scaling are both data reduction techniques and may not accurately represent the true proximity of key topics. In Dendrograms, while keywords that co-occur or cases that are similar tend to appear near each other, the sequence of keywords do not represent a linear representation of those distances as Dendrograms only specify the temporal order of the branching sequence. Therefore, researchers construct proximity plots, reported to be the most accurate way to graphically represent the distance between key topics (14). Proximity plots are not a data reduction technique, but rather a robust visualization tool that extracts information from huge amount of data stored in distance matrices (14).

After producing granular taxonomies of key concepts (Figure 1), researchers performed link analysis using network graphs to visualize the connections (and the strengths of direct and indirect associations) between key CAV Big Data concepts. Fuzzy logic and fuzzy c-means algorithms are used to achieve this (15). After link analysis, the overarching goal is to develop a CAV Big Data logical architecture for summarizing the current body of knowledge and spotting exciting future opportunities pertaining to implications of Big Data for safety monitoring and assessment (1). Throughout the project, STC consortium partners actively participated in this study by providing their insights and input.

Table 1 provides an overview of the tasks completed in this research project, as well as the procedures and key outcomes in the context of CAV Big Data for each task. For example, the "Frequency/Content Analysis" task involves the application of univariate frequency analysis on words or word categories to update an inclusion dictionary developed in a previous state. The results include a series of descriptive statistics, as well as bar and pie charts.

**Table 1: Methods and Outcomes**

| Task | Procedure | Key outcome |
|---|---|---|
| Stemming & categorization process | • Application of natural language processing routines to reduce inflected words to common stem or root form.<br>• Categorizing specific words, word patterns, or phrases to a unique category, e.g. words such as “good”, “excellent”, or “satisfied” may be coded as instances of “positive evaluation”. | Development of detailed specified inclusion dictionary related to CAV Big Data research |
| Frequency/content analysis | • Performing univariate frequency analysis on words or categories & updating the inclusion dictionary developed in previous task | Descriptive statistics, bar charts, pie charts |
| Concept/entity extraction | • Application of machine learning algorithms & statistical procedures to uncover hidden thematic structure of text collection<br>• Using factor analysis for concept/topic extraction | Most important topics, phrases, & Big Data concepts, categorization dictionary |
| Statistical analysis & visualizations | • Conducting co-occurrence analysis for each specific concept/topic i.e. hierarchical cluster analysis & multidimensional scaling<br>• Correspondence & multivariate analysis to quantify high dimensional correlations<br>• Automatically updating categorization dictionary | Correlations (within and between categories), Dendrograms, 2D/3D concept maps, & proximity plots |
| Link analysis | • Application of fuzzy logic and fuzzy c-means algorithms to generate high dimensional network graphs for each Big Data CAV concept | Big Data CAV logical architecture |

## 2. Conceptual Framework for Identifying and Categorizing CAV Big Data Research

The main objective of this research is to use advanced data mining and text analysis tools to synthesize studies, papers, and reports related to CAV data utilized for safety improvements and extract key topics, patterns, and trends. Figure 2 illustrates the conceptual framework of this research. Big data used for safety studies contain both conventional and new data sources. Conventional data sources include Roadway Parameters, Roadway Elements, Land Use, and Planning. Additionally, new technologies, sensors, and data collection methods have allowed researchers to access new data sources. These new data are mostly collected automatically and continuously in real time and therefore require large amounts of storage, making them Big Data sources. Data collected from connected and automated vehicles are one of the most important data sources available today, providing researchers with useful information regarding trajectory, component status, and other relevant travel information. Although studies which have used Big Data peruse a broad range of objectives and concepts such as mobility, security, accessibility, and sustainability, this research is mainly focused on safety and infrastructure design studies.

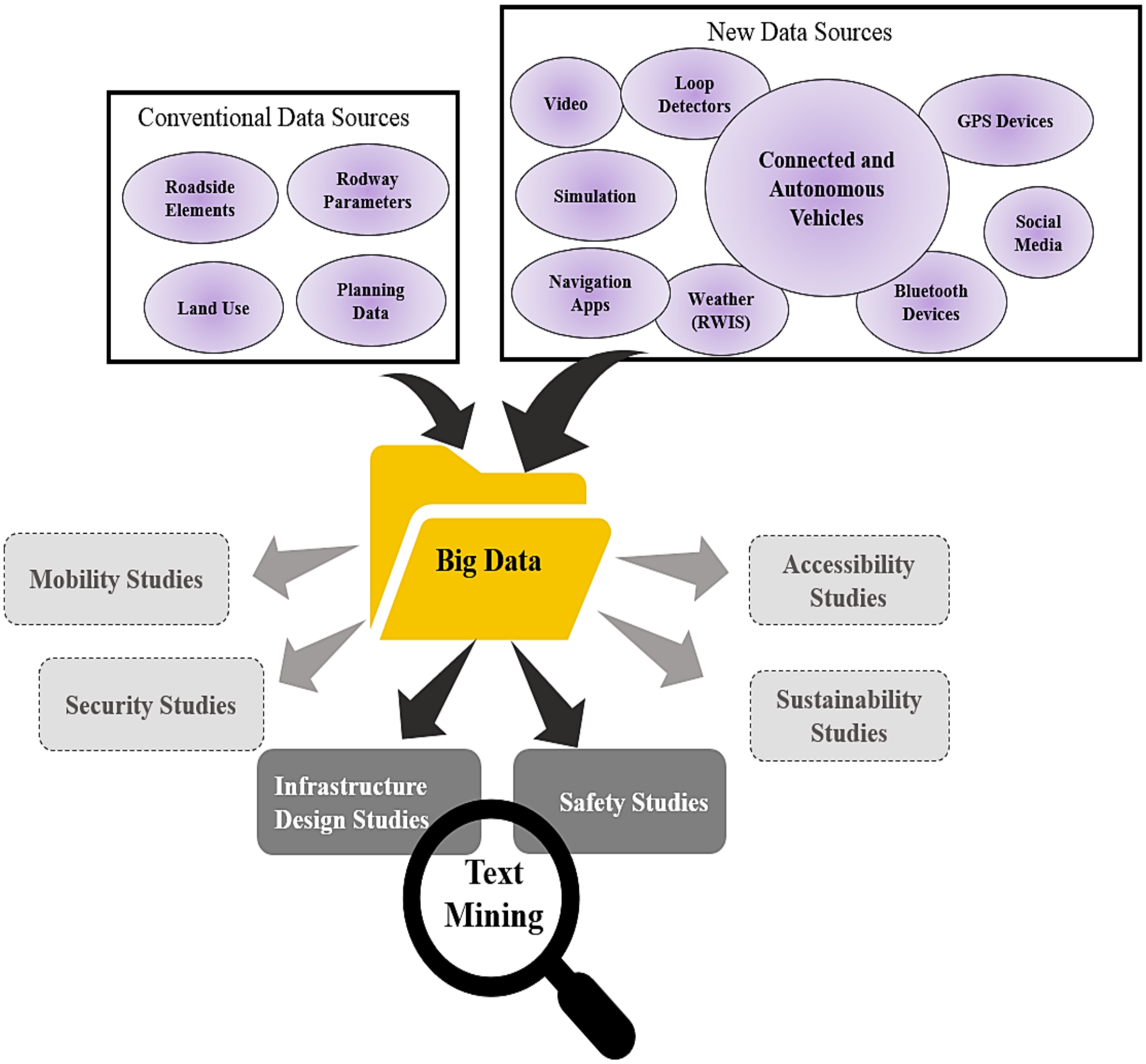

**Figure 2: Conceptual Framework for On-going CAV Big Data Research**

## 3. Identifying Safety-Enhancing Data Needs for Connected and Automated Vehicles

The safety-enhancing data of connected and automated vehicles can be categorized into two groups: conventional data sources and new data sources. Conventional data sources are land use, roadside elements, roadway parameters, planning data, loop detectors, videos, Global Positioning Systems (GPS), Bluetooth devices, social media, and weather. The rapid growth of vehicle usage, vehicle connectivity, infrastructure systems, and travelers has extended large-scale data sources to include stationary radars, lidars, in-vehicle devices, and connected and automated vehicles. The U.S. Department of Transportation (USDOT) supports pilot deployment programs that advance connected and automated technologies, focusing on mobility and safety applications. Three of these programs are located in New York City, Wyoming, and Tampa and they implement connected and automated technologies to assess their real-world potential to increase safety, enhance mobility, and reduce environmental impacts.

Connected Vehicle Pilots Deployment Programs are large-scale sources of Basic Safety Messages (BSMs) that integrate mobile devices, connected vehicles, and infrastructure to measure the benefits of these technologies and resolve existing connected vehicle deployment issues. The focus of the New York City pilot program is safety (i.e., reduce crash related deaths, injuries, and property damage). To achieve an increase in safety, the program is investing in Dedicated Short-Range Communication (DSRC) technology, vehicle-to-vehicle (V2V), vehicle-to-infrastructure (V2I), and infrastructure-to-pedestrian (V2P) communications. The Wyoming pilot program on Interstate 80 (I-80), which is a major corridor of freight movement, plans to implement V2V and V2I communication systems using DSRC in order to assess and improve safety and mobility in this corridor. The Tampa-Hillsborough Expressway pilot program benefits from wireless communication between vehicles crosswalks, and traffic signals to enhance safety in a congested urban area. Furthermore, the National Research Energy Lab (NREL) provides large-scale, detailed transportation data from travel surveys including GPS information, vehicle characteristics, and traveler demographics through the Transportation Secure Data Center (TSDC). These data sources can be applied to a variety of research endeavors, including travel demand modeling, congestion mitigation, and validation of other transportation data sources. Finally, the National Household Travel Survey (NHTS) conducted by the Federal Highway Administration (FHWA) is the only host of national data including all-mode non-commercial trips of travelers, their households, and vehicles.

### 3.1. A Comprehensive Database of Nationwide and International Big Data Research

This study created a comprehensive database of nationwide and international peer-reviewed high-quality research work. The real-world application of Big Data sources for safety monitoring, assessment, and improvement are the specific focus. Researchers accessed relevant studies from the US-DOT Data Capture and Management program, nationwide Big Data repositories maintained by the Transportation Research Board, and the European Big Data and B2B (Business-to-Business) digital platforms initiative. The efforts reflected in these reports are divided into two main parts: (i) transportation safety and (ii) infrastructure design in connected and automated vehicles (CAVs). These efforts include a comprehensive search that collects papers revolving around CAV safety with a special focus on Big Data. The study utilizes different search engines such as Google Scholar, Science Direct, and Web of Science to select the intended studies. Example keywords include "connected vehicles", "automated vehicles", "big data", "safety", "crash." Researchers also surveyed the references of each literature in order to find and add relevant sources to the main database. In the end, the database included up to 100 documents comprised of journal papers, conference papers, theses, and reports. To catalogue the literature relevant to infrastructure design, the survey used keywords such as "CAV infrastructure design", "CAV infrastructure architecture", and "CAV infrastructure requirement." Infrastructure design included a total of 10 documents including technical reports, journal papers, and white papers

### 3.2. Literature Review

While self-driving cars have positive impacts on mobility (16), such as increasing mixed-traffic performance and capacity (*17; 18*), increasing the efficiency and throughput of intersections (19), and lowering fuel consumption (20), connected and automated vehicle deployment offers both positive and negative impacts on safety. CAVs have the potential to reduce conflicts (*21; 22*) by improving the safety of lane change maneuvers (23-26) through advanced technologies, such as cooperative adaptive cruise control (27), driving assistance systems (*28; 29*), "dynamic-lane changing trajectory planning model" (25), applications using dedicated short range communication (DSRC) (30), Internet of Cars (31), blind spot monitoring (32), lane departure warning, and forward collision warning (33). Pek et al. found that self-driving vehicles can increase the safe lane change from 39.83 to 56.7 percent (34). It is believed that queue warning systems (35), collision notification systems (*36; 37*), and cooperative adaptive cruise control (CACC) (38-42) improve the safety conditions of the network by reducing the number of rear-end conflicts. In addition, studies show that "inter-vehicle warning information system" and "forward collision warning and autonomous braking systems" reduce the number of rear-end conflicts (*43; 44*) - up to approximately 84.3 percent for inter-vehicle warning systems (45); 3.2 percent for forward collision warning systems (46), and 7.7 percent for autonomous braking systems (*46; 47*).

Data from connected vehicles have the potential to advance real-time road safety management (48) especially by identifying the safety-critical events at high-risk locations (*5; 49-56*). Data from connected vehicles gives the opportunity for researchers to analyze volatility in micro-level driving behavior (*50; 57-59*). In collision prone traffic conditions, the interaction aware model enhances the prediction of collision probability up to 10 percent for autonomous vehicles (60). Li et al. valued the safety benefits of CAVs for Americans by up to $76 billion each year, which reflects the crash related cost (61). For the sake of connected vehicle technologies and low-level automation systems, the number of accidents in major categories can be reduced (*62*) by perceiving and avoiding the unsafe events (*63; 64*). Safety of CAV navigation can be achieved through V2V applications (*65*), inter-vehicle communication (*66; 67*), and "Cooperative Intelligent Driver Model" (*68*). This reduction value could be 1.77 accidents per 1000 drivers (*69*). One study shows that intelligent cruise control systems (ICCS) can reduce accidents by up to 7.5 percent (*70*). This reduction would be 32.99 to 40.88 percent for light and heavy vehicles, respectively, while 35 percent reduction in near-crash events can be achieved in adverse weather condition (*71*). Thus, crash reductions due to increasing the automated vehicle market can decrease the risk of secondary crashes by about 10 percent, decrease crash related delays and increase reliability (*72-76*). Moreover, connected and automated vehicle platooning can improve

passenger comfort (*77; 78*), as well as longitudinal safety (*79-82*). Also, CAV platooning can positively impact safety for heavy duty vehicles (*83*).

In contrast, it has long been debated that CAVs will not only increase vehicle miles traveled by at least 13 percent in a region (*84*), but will also have negative impacts on safety that include technology failures (*85; 86*), privacy violations (*87; 88*) and cyber-attacks (*89; 90*). Targeting the machine vision, Global Positioning Systems (GPS), in-vehicle devices, radar, lidar, maps, acoustic sensors, infrastructure equipment (*91*) are examples of CAV safety issues. This has prompted many researchers to propose platforms of safety impact analysis (*92; 93*) benefiting vehicle dynamics, sensing errors, and crash severity quantifications (*94*). However, safety forecasts for highly and fully automated vehicles are mostly based on assumptions that still need to be refined and validated in a more detailed fashion (*95; 96*). Though automated technologies can reduce the drivers' workload and help for safer control (*97*) and less (human) error (*98-100*), it has been found that automated vehicles (AV) can cause passive fatigue and decrease the alertness of drivers, which in turn slows manual takeover (*101-105*). In addition, drivers' reaction time in disengagement situations increases with increasing vehicle mile traveled in CAV's, which undermine the safe transition mode (*106; 107*). Nevertheless, different auditory systems employed in AVs can reduce the "reflecting information processing time" as a part of reaction time; however, the auditory systems do not change the "reflecting allocation of attention" part of reaction time (*108*).

Higher penetration rates of CAV use requires a major revision in safety standards (*109*). With the emergence of high-level automated vehicles, the responsibility of driving errors moves from drivers to vehicles and vehicle manufacturers, but current standards and regulations, such as ISO 26262, which is an international standard for functional safety of electrical and/or electronic systems in production automobiles defined by the International Organization for Standardization in 2011, do not address this issue (*110*). The impacts of CAVs in terms of interactions with other components of the transportation system also remain uncertain (*87*) and trust in AV technology is relatively low in interactions with pedestrians (*111; 112*).

Studies have reviewed all key aspects of the emerging of CAVs, their impacts, and requirements (*113*); in this report, researchers focus on deriving high-quality structured information from safety-oriented Big Data studies. This helps researchers in systematically structuring and creating a synopsis of the current body of Big Data knowledge. Researchers extract meaningful information and produce new insights by spotting Big Data innovation indicators. Figure 3 demonstrates the framework for identifying mechanisms that may affect harm cost through an investigation of both negative and positive impacts of CAV safety variables. To this end, the title, abstract and result of each document are inspected to investigate topics of interest and categorize these topics based on their positive and negative impacts of CAVs on harm cost. For example, conflict reductions (i.e. safe lane change maneuvers, safe merging-in and rear-end collision reduction) result in harm cost reduction, while cyber-attacks and privacy violations result in harm cost increase.

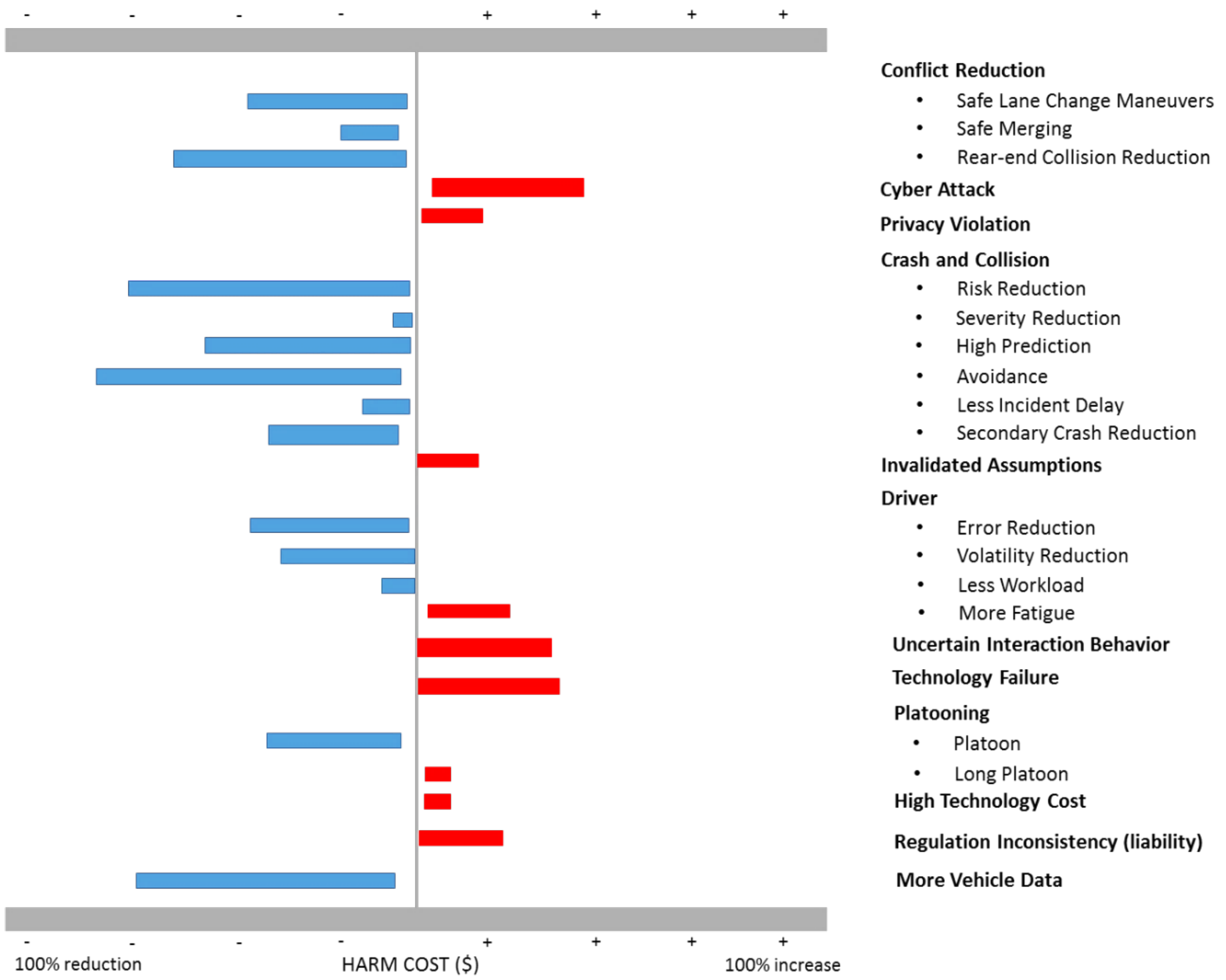

**Figure 3: Framework for Mechanisms that May Affect Harm Cost**

### 3.3. Big Data Related Work Classification

To synthesize the CAV Big Data innovation process, Big Data generation was classified based on three different perspectives: 1) Physical classification (vehicles, V2V and V2I communications, DSRC, LTE, and transportation management centers), 2) Methodological classification (geographic information systems (GIS), algorithms and simulation, modeling and fusion, visualization), and 3) Institutional classification (organizational structures, incentives, and initiatives). Figure 4 illustrates the classification of Big Data generation based on the three categories. The values in the figure present the number of studies in which each technology, methodology, or initiatives are used for Big Data generation. For example, researchers reviewed 43 resources in the physical category; 48 resources in the methodological category and none in the institutional category, indicating a clear gap in research.

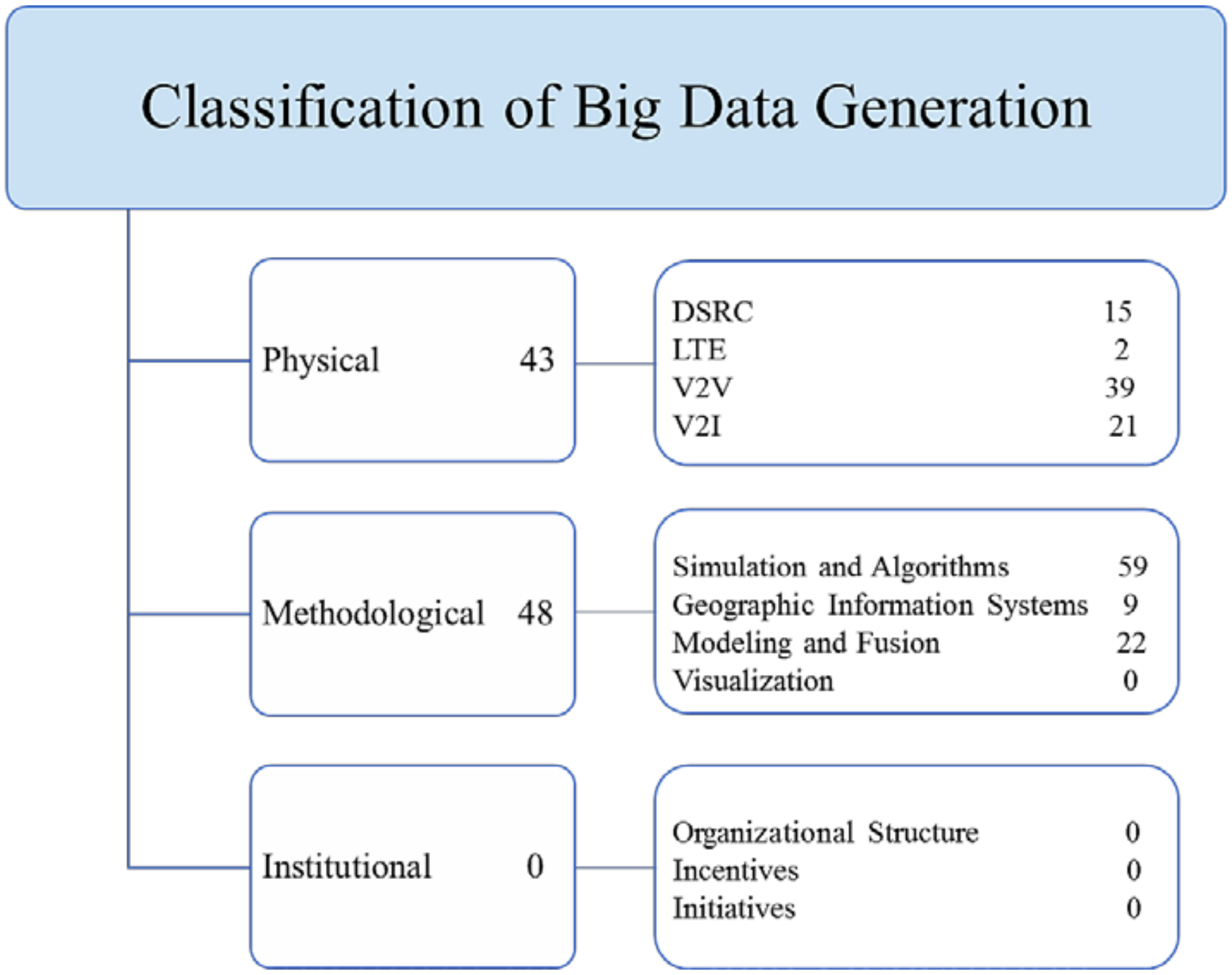

**Figure 4: Classification of Big Data Generation**

## 4. Frequency/Content Analysis and Concept/Entity Extraction

To perform content analysis, an "inclusion dictionary" was developed. To achieve this goal, frequency analysis was applied to identify the most frequently used keyword and phrases in the literature. After performing an initial analysis, an exclusion list was made in order to remove the words that either carry little semantic value, such as propositions, conjunctions, etc., or those frequently-used words with little discriminative value. The "inclusion dictionary" is developed to merge different forms of a word (e.g. vehicles and vehicle) to consider them as a single word. Then, all words or phrases that do not meet a minimum frequency are removed from the final list. Table 2 shows the word frequency and descriptive statistics of the final word list.

The columns of "Frequency" and "% Cases" represent the number of occurrences of the keyword or phrase and percentage of cases where the keyword appears respectively. In the same way, the column "TF*IDF" represents the term frequency multiplied by inverse document frequency. The underlying assumption is that when a term appears in a document more frequently, it is more likely to be a representative of the content. However, the more documents the term appears in, the less discriminating it would be (*114*). Appearing 2474 times in 78 percent of the collection, "crash" has the highest frequency in the reviewed literature. The next most frequent word is "lane" used in 92 percent of the articles. Similarly, Speed, Automated, AV, Collision, Intersection, Brake, Accident, Acceleration, and Risk are among the most repeatedly used words in studies.

Figure 5 presents the distribution graphs of keywords based on the frequency and percent of studies they were appeared in respectively. For example, the figure visually highlights the frequency difference between the word "crash" and "lane". Figure 6 illustrates the distribution or frequency of words based on the percentage of cases, where total cases equals 100. In this graph, however, the word "lane" has a higher percentage of cases than the word "crash" and while one of the least frequent words "signal" appears 330 number of times, it appears in nearly 60 percent of all cases. Finally, Figure 7 demonstrates the word cloud plot based on the frequency statistics of the word list. In a word cloud, frequencies are converted to words with different sizes, the more frequently the word appeared in the studies, the larger the word would be in the plot. For example, the word cloud emphasizes the words Crash, Speed, and Lane, which are the top three words listed in Figure 5.

**Table 2: Keyword Frequency and Descriptive Statistics (N =100 studies)**

| Keyword | Frequency | % Cases | TF • IDF | Keyword | Frequency | % Cases | TF • IDF |
|---|---|---|---|---|---|---|---|
| Crash | 2474 | 78.00% | 267.0 | Fatigue | 461 | 16.00% | 366.9 |
| Lane | 2245 | 92.00% | 81.3 | Benefit | 456 | 49.00% | 141.3 |
| Automate | 1731 | 79.00% | 177.2 | Velocity | 453 | 44.00% | 161.5 |
| AC | 1658 | 25.00% | 998.2 | Platoon | 451 | 27.00% | 256.5 |
| Collision | 1341 | 85.00% | 94.6 | Cooperation | 445 | 50.00% | 134.0 |
| Intersection | 1255 | 54.00% | 335.8 | Penetration | 445 | 41.00% | 172.3 |
| Brake | 1094 | 80.00% | 106.0 | Standard | 440 | 70.00% | 68.2 |
| Warning | 1089 | 61.00% | 233.8 | Rear | 434 | 55.00% | 112.7 |
| Accident | 1035 | 79.00% | 106.0 | CACC | 418 | 19.00% | 301.5 |
| Acceleration | 984 | 66.00% | 177.6 | Longitudinal | 416 | 53.00% | 114.7 |
| Risk | 972 | 67.00% | 169.1 | Response | 413 | 59.00% | 94.6 |
| Communication | 946 | 79.00% | 96.8 | Attack | 408 | 13.00% | 361.5 |
| Simulation | 936 | 77.00% | 106.2 | Cost | 400 | 52.00% | 113.6 |
| Autonomous | 902 | 64.00% | 174.8 | Pedestrian | 393 | 31.00% | 199.9 |
| Automation | 887 | 51.00% | 259.4 | Dynamic | 384 | 58.00% | 90.8 |
| CAV | 765 | 15.00% | 630.3 | Maneuver | 377 | 45.00% | 130.7 |
| Behavior | 744 | 76.00% | 88.7 | Steering | 374 | 53.00% | 103.1 |
| CV | 737 | 21.00% | 499.5 | Message | 364 | 48.00% | 116.0 |
| Algorithm | 630 | 66.00% | 113.7 | Assistance | 360 | 55.00% | 93.5 |
| Task | 604 | 54.00% | 161.6 | Market | 359 | 51.00% | 105.0 |
| Reaction | 589 | 59.00% | 135.0 | Avoidance | 356 | 49.00% | 110.3 |
| Decision | 575 | 70.00% | 89.1 | Infrastructure | 345 | 62.00% | 71.6 |
| Safe | 575 | 71.00% | 85.5 | Lateral | 342 | 49.00% | 106.0 |
| Trajectory | 571 | 48.00% | 182.0 | Headway | 337 | 42.00% | 127.0 |
| Deceleration | 566 | 55.00% | 147.0 | Signal | 330 | 59.00% | 75.6 |
| ACC | 516 | 40.00% | 205.3 | Turn | 330 | 56.00% | 83.1 |
| Network | 512 | 61.00% | 109.9 | Gap | 327 | 55.00% | 84.9 |
| Volatility | 508 | 7.00% | 586.7 | Conflict | 326 | 31.00% | 165.8 |
| Sensor | 495 | 67.00% | 86.1 | Threshold | 309 | 42.00% | 116.4 |
| TTC | 474 | 28.00% | 262.0 | | | | |

NOTES: TF • IDF = Term frequency multiplied by inverse document frequency (used for ranking relevance), CAV=Connected and Automated Vehicles, CV=Connected Vehicles, ACC=Adaptive Cruise Control CACC=Cooperative Adaptive Cruise Control, TT=Time to Collision.

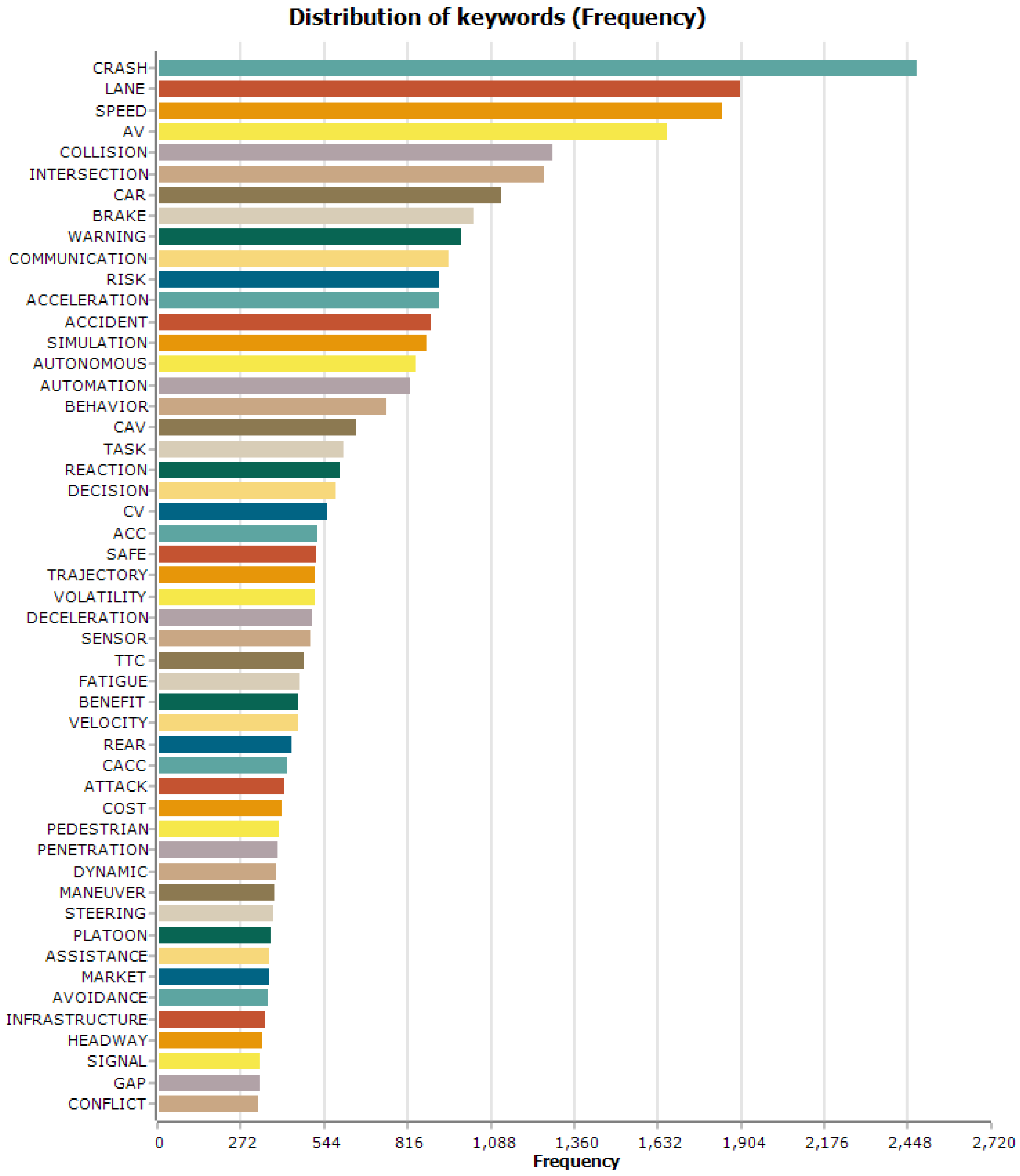

**Figure 5: Distribution of Words (Frequency)**

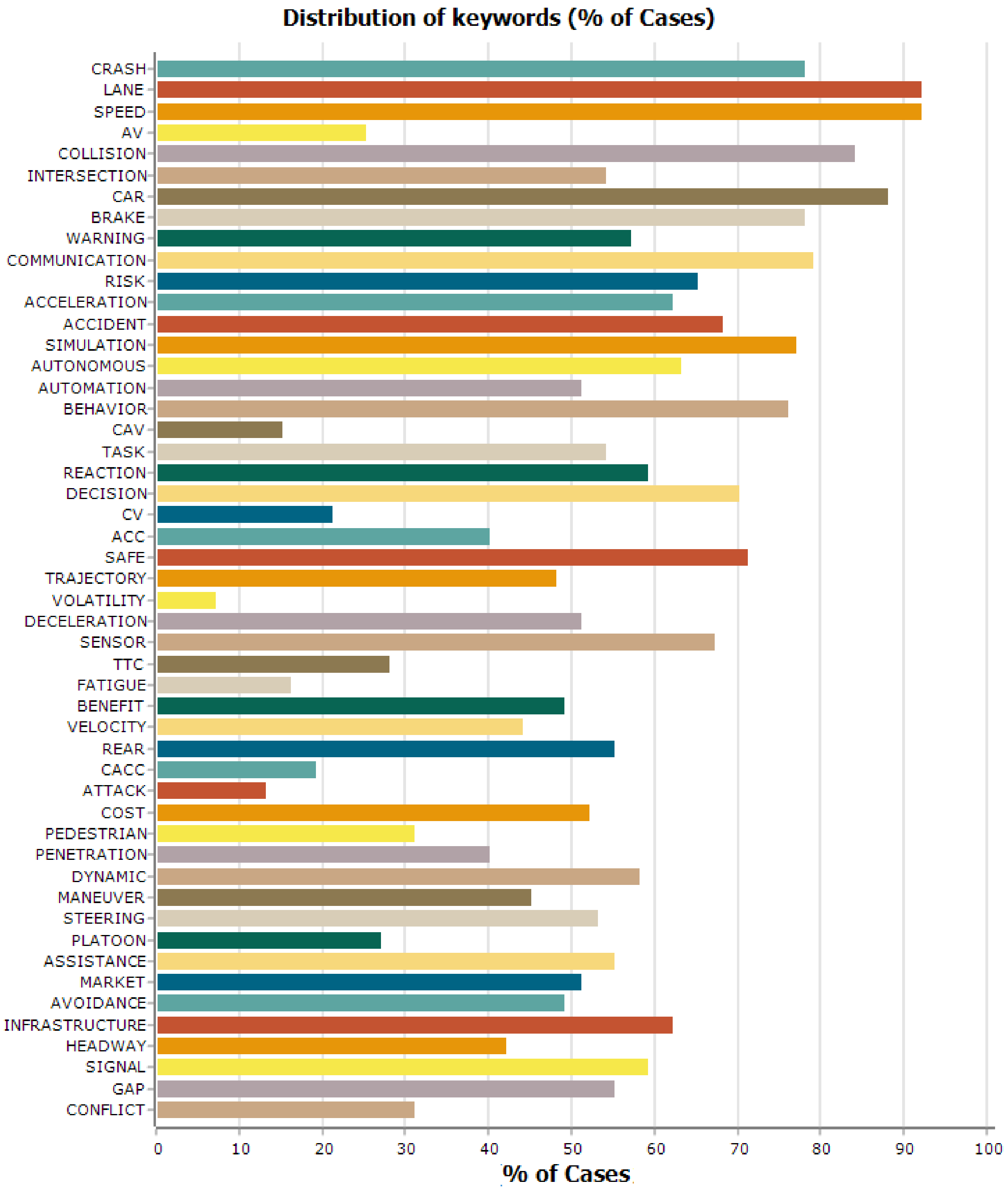

**Figure 6: Distribution of Words (% of cases)**

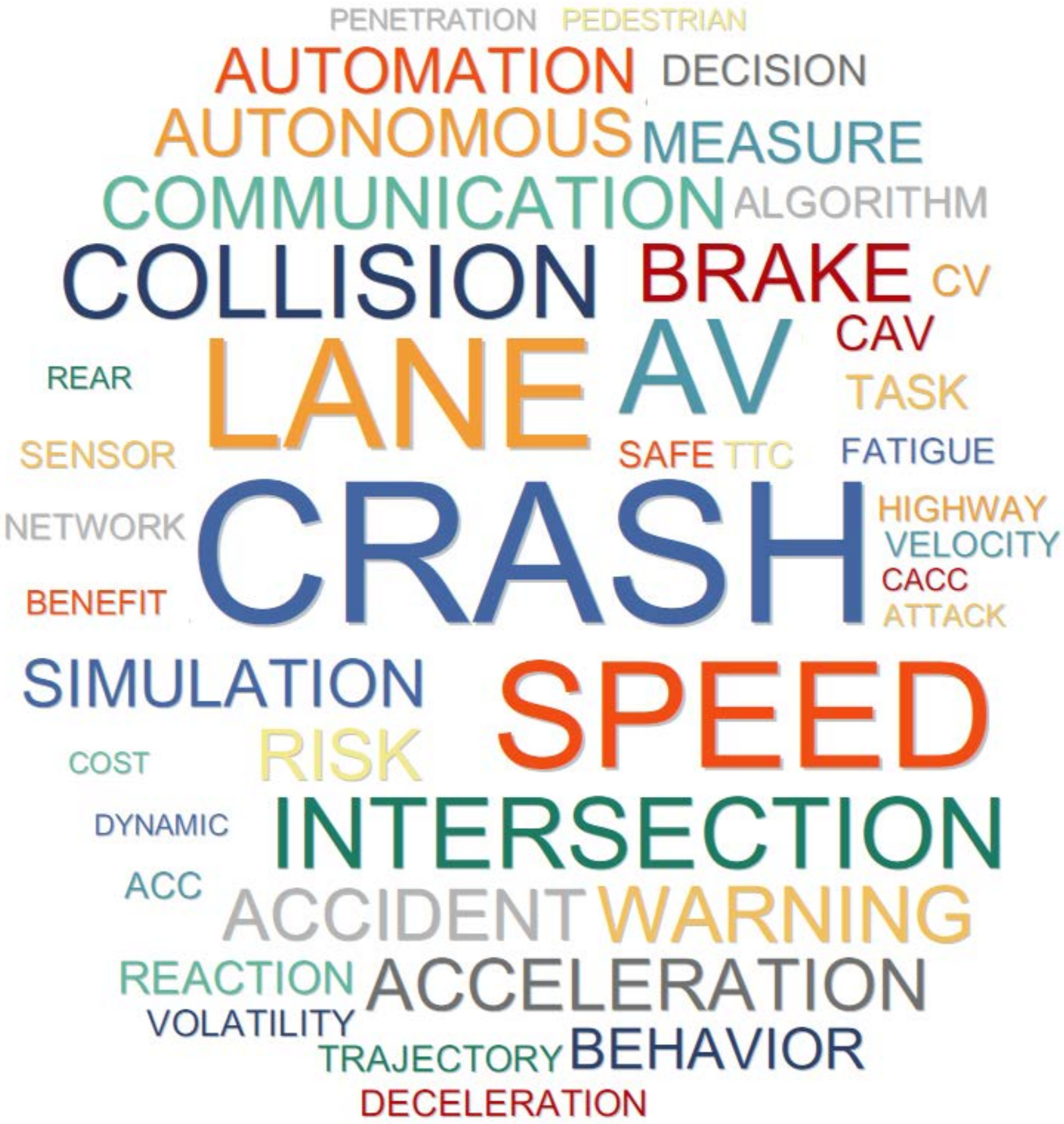

**Figure 7: Word Cloud of High Frequency Words**

Similar to the frequency analysis of keywords, an analysis was performed on phrases to identify the most frequent groups of words in the collected studies. Table 3 shows the results for the phrase frequency and descriptive statistics. The results show that “Lane Change” is the most frequently used phrase in the literature appearing 825 times in 63 percent of studies. “Reaction Time” has the second highest frequency appeared 452 times in 47 percent of the articles. Likewise, “Rear End”, “Car Follow”, “Cruise Control”, and “Market Penetration” are other phrases with high frequencies in the collected studies. Appearing only 33 times are the phrases "Trajectory Data" at 19 percent and "Data Collect" at 22 percent.

Figure 8 presents the distribution of the frequency and the percent of phrases that appeared in studies. The phrase "Lane Change" is mentioned almost twice as frequently as "Reaction Time", while "Fuel consumption" and "Rear End Collision Risk" are mentioned about 50 times. Figure 9 illustrates the distribution of most frequently used phrases based on percentage of use. In this graph, the phrase "Lane Change" is used most frequently, followed in percentage of frequency by "Cruise Control", "Adaptive Cruise Control", "Reaction Time", "Real World" and "Real Time". Finally, Figure 10 shows the phrase cloud for frequency statistics of the selected phrase list at a glance. Most obvious is the phrase “Lane Change”.

**Table 3: Phrase Frequency and Descriptive Statistics**

| Phrase | Freq | % Cases | TF • IDF | Phrase | Freq | % Cases | TF • IDF |
|---|---|---|---|---|---|---|---|
| Lane Change | 825 | 63.00% | 165.5 | Cyber Attack | 83 | 6.00% | 101.4 |
| Reaction Time | 452 | 47.00% | 148.2 | Passenger Comfort | 83 | 3.00% | 126.4 |
| Rear End | 370 | 39.00% | 151.3 | Decision Make | 81 | 35.00% | 36.9 |
| Car Follow | 241 | 38.00% | 101.3 | Surrogate Safety Measure | 75 | 15.00% | 61.8 |
| Cruise Control | 227 | 54.00% | 60.7 | CACC System | 73 | 5.00% | 95.0 |
| Market Penetration | 204 | 27.00% | 116.0 | Road Condition | 73 | 26.00% | 42.7 |
| Real World | 171 | 45.00% | 59.3 | Lane Change Trajectory | 70 | 2.00% | 118.9 |
| Rear End Collision | 169 | 20.00% | 118.1 | Warning Index | 70 | 3.00% | 106.6 |
| Steering Wheel | 167 | 27.00% | 95.0 | Safety Performance | 65 | 22.00% | 42.7 |
| Crash Frequency | 160 | 9.00% | 167.3 | Wireless Communication | 63 | 21.00% | 42.7 |
| Dynamic Intersect. | 156 | 1.00% | 312.0 | Time Headway | 61 | 17.00% | 46.9 |
| Adaptive Cruise Ctl | 154 | 48.00% | 49.1 | Safe Distance | 60 | 16.00% | 47.8 |
| Collision Avoidance | 153 | 29.00% | 82.3 | Longitudinal Acceleration | 58 | 13.00% | 51.4 |
| Accident Analysis and Prevention | 149 | 30.00% | 77.9 | Basic Safety Message | 55 | 15.00% | 45.3 |
| Assistance System | 142 | 32.00% | 70.3 | Rear-End Collision Risk | 55 | 5.00% | 71.6 |
| Safety Benefit | 135 | 22.00% | 88.8 | Fuel Consumption | 54 | 17.00% | 41.6 |
| Real Time Data | 132 | 45.00% | 45.8 | Highway Capacity | 54 | 6.00% | 66.0 |
| Signalized Intersection | 130 | 12.00% | 119.7 | Random Parameter Poisson | 49 | 4.00% | 68.5 |
| Crash Avoidance | 121 | 19.00% | 87.3 | Crash Type | 48 | 15.00% | 39.5 |
| Collision Risk | 119 | 13.00% | 105.4 | Lane Departure Warning | 48 | 19.00% | 34.6 |
| Crash Scenario | 114 | 8.00% | 125.0 | Connect Vehicle Data | 42 | 7.00% | 48.5 |
| Collision Warning | 110 | 33.00% | 53.0 | Connect Vehicle Environment | 40 | 16.00% | 31.8 |
| Surrogate Safety | 107 | 20.00% | 74.8 | Crash Avoid. Tech. | 40 | 4.00% | 55.9 |
| Time to Collision | 105 | 34.00% | 49.2 | CV Technology | 38 | 15.00% | 31.3 |
| Warning System | 102 | 32.00% | 50.5 | Big Data | 37 | 9.00% | 38.7 |
| Lateral Accelerate | 100 | 15.00% | 82.4 | Safety Pilot | 35 | 10.00% | 35.0 |
| Crash Data | 94 | 16.00% | 74.8 | System Failure | 34 | 12.00% | 31.3 |
| Crash Risk | 92 | 20.00% | 64.3 | Time to Collision | 34 | 21.00% | 23.0 |
| Rear End Crash | 90 | 23.00% | 57.4 | Data Collect | 33 | 22.00% | 21.7 |
| Adapt. CC System | 89 | 11.00% | 85.3 | Trajectory Data | 33 | 19.00% | 23.8 |
| Safety Impact | 86 | 27.00% | 48.9 | | | | |

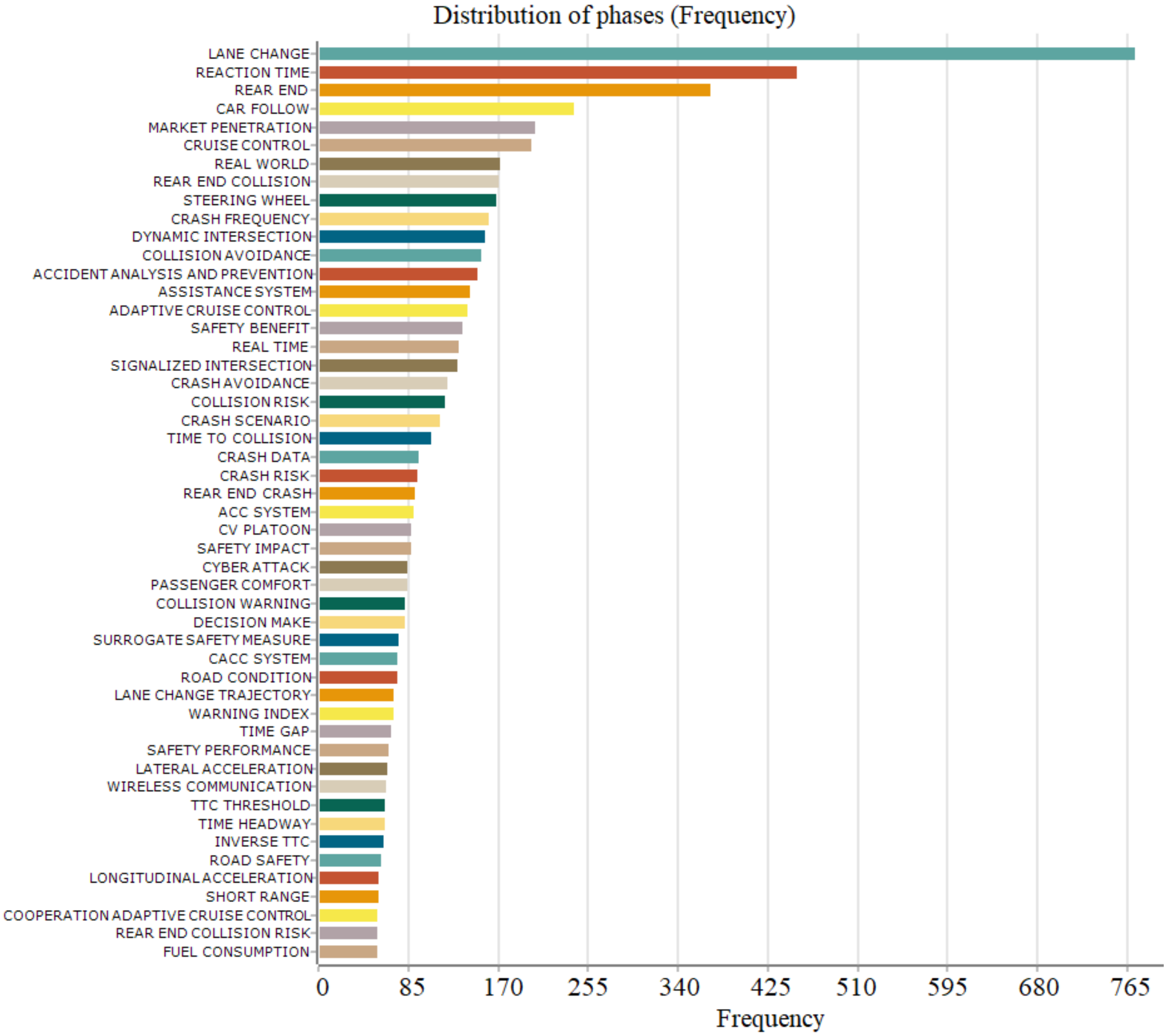

**Figure 8: Distribution of Most Frequent Phrases (Frequency)**

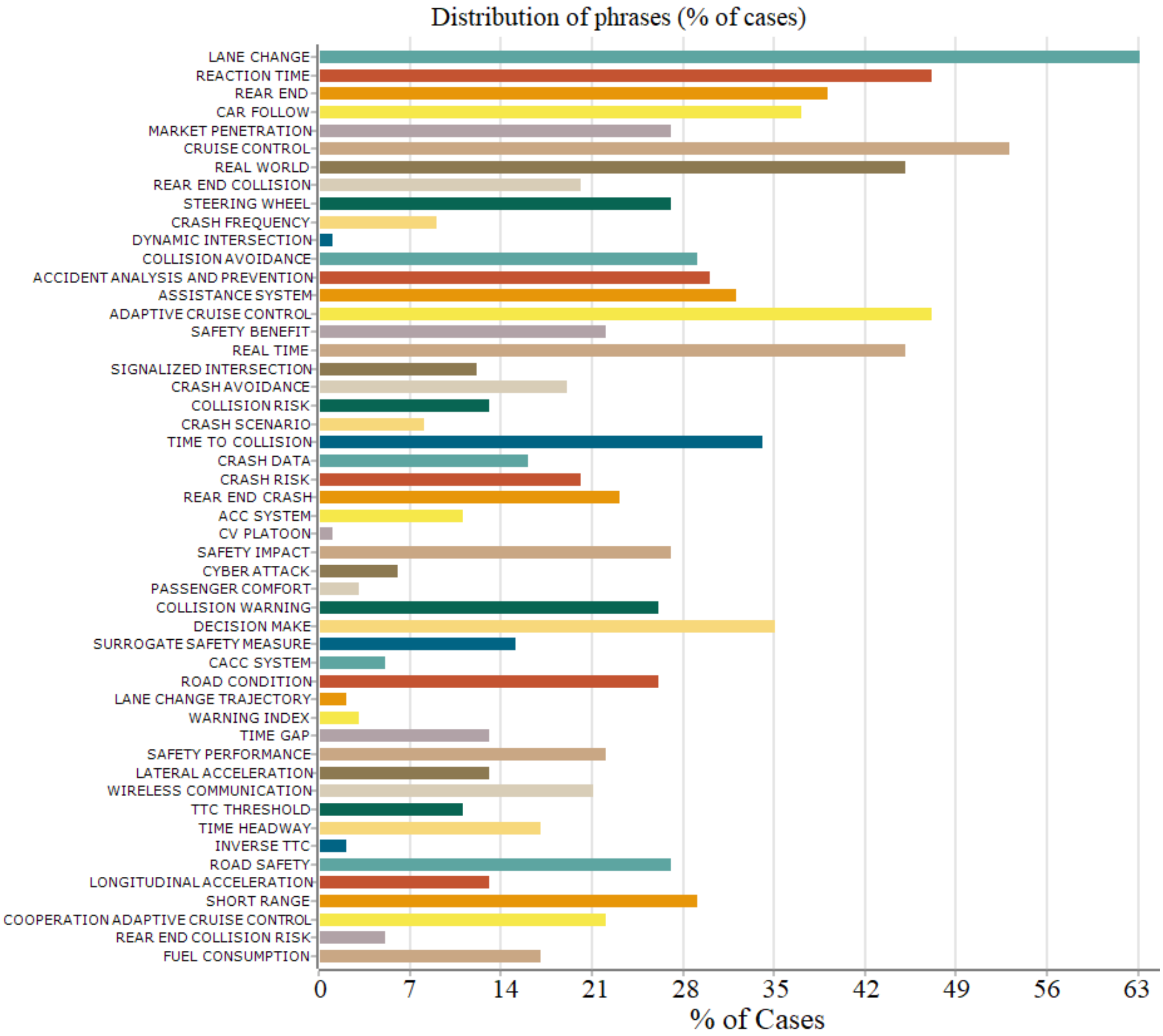

**Figure 9: Distribution of the Most Frequent Phrases (% of cases)**

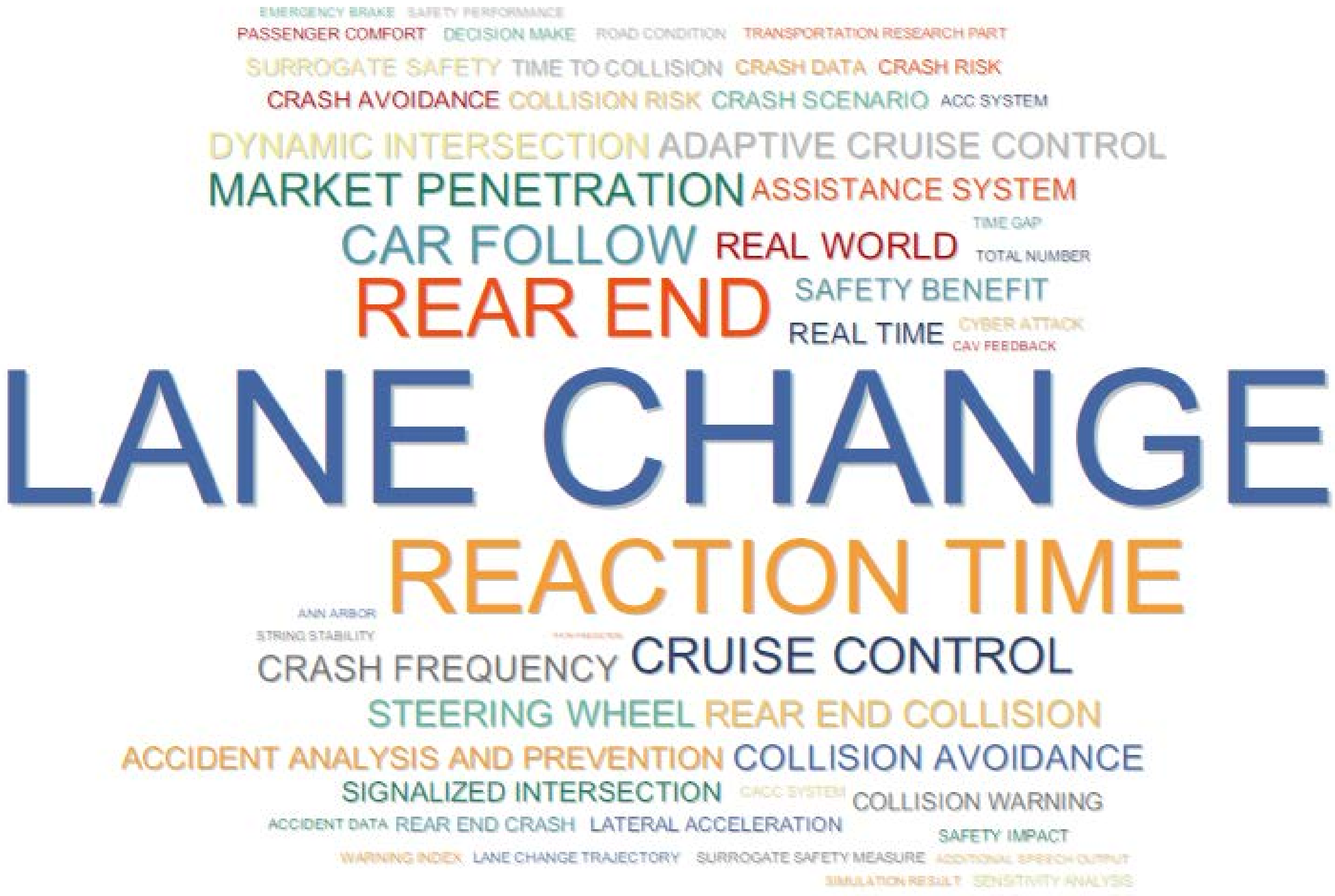

**Figure 10: Phrase Cloud of the Most Frequent Phrases**

Table 4 shows the results of topic extraction using the Factor Analysis method. Generally, higher eigenvalues and the coherence indicate higher variability explained by the topic. This means the topic is more important with more coverage. Therefore, eigenvalues and coherence are good measures to determine which topics are most striking (*115*). Having the highest eigenvalue with a coherence of 0.442, the topic of "Market Penetration Impacts" appears 157 times in 70 percent of the studies. The second highest eigenvalue belongs to "Autonomous Lane Changing" with an eigenvalue of 2.53 and a coherence of 0.426, which appeared 442 times in 92 percent of the studies. Other topics of interest based on decreasing eigenvalues include: "Collision Avoidance and Warning", "Diving Volatility", "Collision Control", "Communication Systems and Data", "Driver Behavior Big Data", "Driving Task, Engagement/ Disengagement", and "Intelligent Transport Systems".

Figure 11 presents word clouds of the topics extracted. In addition to Qualitative Data Analysis software, the open source statistical software "R" was used for topic modeling and visualization. The "topicmodels" and the "wordcloud" package were used for Latent Dirichlet Allocation (LDA) and visualization of the topics and their keywords (*116*). Figure 11 also illustrates the word clouds of the extracted topics and their keywords. These word clouds provide a better understanding through visualization of important topics and the keywords around them. For example, where Autonomous Lane Change is the topic of study, safety, accident, and vehicle trajectories are among the most important subtopics associated with Autonomous Lane Change. Likewise, studies carried out on Driving Volatility mostly dealt with connected vehicles, crash risk, and safety in intersections.

## Table 4: Results of Topics Extraction through Factor Analysis

| Topic | Keywords | Coherence | Eigen value | Freq | % Cases |
|---|---|---|---|---|---|
| Market Penetration Impacts | Penetration; Market; Technology; CV; Simulation; Impact; Safety; Market Penetration; Penetration Rates; | 0.442 | 6.44 | 157 | 70% |
| Collision Avoidance and Warning | Warning; Avoidance; Collision; Benefit; Brake; Technology; Crash; Collision Avoidance; Warning; | 0.394 | 2.07 | 180 | 75% |
| Collision Control | Collision; Rear; End; Risk; Control; Connected; End Collision; Collision Risk; End Crashes; Moving Rear; Control system; | 0.405 | 1.80 | 218 | 56% |
| Communication Systems and Data | Communication; Information; Cruise; Control; Simulation; Speed; Infrastructure; Performance; Data Cooperative Adaptive Cruise Control; Communication Technology; | 0.390 | 1.64 | 175 | 75% |
| Autonomous Lane Changing | Autonomous; Lane; Accident; Autonomous; Development; Safety; Data; Road; Changing; Safe; Analysis; Av; Autonomous Vehicles; Lane Changing; Safety Benefits; Accidents; | 0.426 | 2.53 | 442 | 92% |
| Driving Task, Engagement/ Disengagement | Engagement; Disengagements; Reaction; Number; Type; Autonomous; Task; Drivers; Reaction Times; Driving Task; Control Of; Autonomous Driving; | 0.378 | 1.46 | 178 | 57% |
| Driving Volatility | Volatility; Intersection; Crash; Speed; Data;<br>Driving Volatility; Crash Frequencies; Intersection Safety; Crash Avoidance; Connected Vehicles; | 0.415 | 2.03 | 256 | 69% |
| Driver Behavior Big Data | Behavior; Decision; Real; Information; Data; Brake; Communication; Time; Warning; Driving Behavior; Lane Changes; Safety Pilot Model Deployment; Connected Vehicles; Big Data; Connected Vehicles; | 0.432 | 1.55 | 211 | 86% |
| Intelligent Transportation Systems | Level; Automation; System; Impact; Performance; Intelligent; Level of Automation; Automated Driving; Intelligent Transportation Systems; | 0.357 | 1.33 | 148 | 75% |

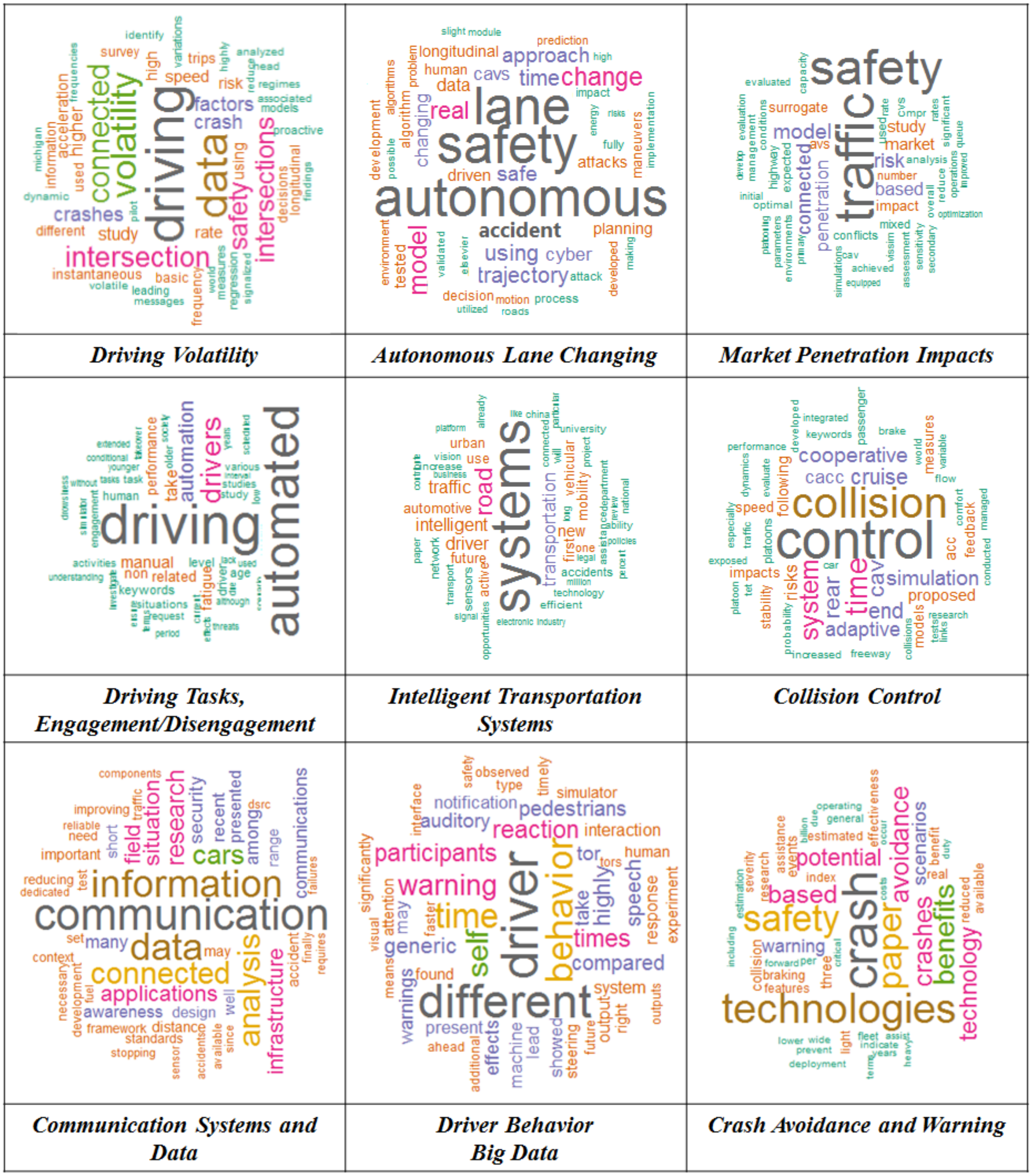

**Figure 11: Key Topics and Related Word Clouds through the LDA Method**

## 5. Statistical Analysis and Visualizations

The application of advanced data analytic techniques in this task will allow researchers to perform hierarchical cluster analysis (HCA) and multidimensional scaling (MDS) of key Big Data concepts and themes. This will help in understanding low dimension co-occurrences of key concepts. Both HCA and MDS are data reduction techniques and are suitable for quantifying low order correlations. For quantifying higher dimension correlations (both within and between categories), correspondence and multivariate analysis will be performed. The results will be visualized using advanced tools such as Dendrograms, 2D/3D concept maps, and proximity plots.

Figure 12 illustrates the concept map of words across key concepts and topics. In this chart, words are plotted with a line between each pair of words showing a strong co-occurrence coefficient. Each cluster of words with a distinct color show a topic or concept on Big Data safety. From the figure, it can be seen that “Driving Task, Engagement Behavior”, “Driving Volatility”, “Communication Technology and System”, “Platooning”, “Crash Avoidance”, and “Lane Change Behavior” are among the most important concepts. Noticeably, there is a strong correlation at the center of the blue cluster belonging to the topic “Driving Task, Engagement Behavior” between some of the most common words in this area of study: Driving, Drivers, Behavior, Task, Automated, Level of automation, Reaction, Breaking, and Steering. It shows that driving behavior and disengagement mostly deal driving task, level of automation, steering and braking systems. Similarly, when it comes to “Communication Technology and System” there is a strong connection between these keywords: System, Communication, Application, Infrastructure, CACC (Cooperative Adaptive Cruise Control), Sensor, GPS, and so on. There are also some links between the words of different clusters showing the correlation between different concepts. For instance, the topic of “Communication Technology and System” is highly correlated with the topic of “Driver Task, Engagement Behavior” indicating that driving engagement behavior significantly is related to the communication technology and systems applied in CAVs.

Proximity plots are an accurate way to visualize keywords that co-occur with one or several target keywords. Figure 13 is the proximity plot of keywords considering Automated Vehicles (AVs) and Connected Vehicles (CVs) as the reference target keywords. This figure presents the major keywords on AVs and CVs and their proximity scores. For example, the keywords of Avoidance, Queue, Platoons, Estimation, Lanes, and Technology have the highest proximity with CVs. Similarly, the keywords of Travel, Government, Testing, Privacy, Autonomous, Pedestrian, and Demand have the highest proximity with AVs. In addition, a comparison can be drawn between the proximity score of the common keyword in AVs and CVs. From the figure, it can be concluded that studies conducted on pedestrian safety and users mostly deal with AVs, and connectivity is not a matter of concern. Likewise, when it comes to market penetration, more studies have been conducted on the adoption and recognition of AVs in the market. Figures 14 and 15 show proximity plots for single target keywords. Figure 14 shows that keywords of Cyber Attack, Capacity, Travel, Penetration, and Platooning have the highest proximity with CAVs. Similarly, Figure 14 indicates that those articles in which different impact of CAVs have been studied, keywords of Penetration, Capacity, Travel, Market, AVs, and Flow have the highest proximities.

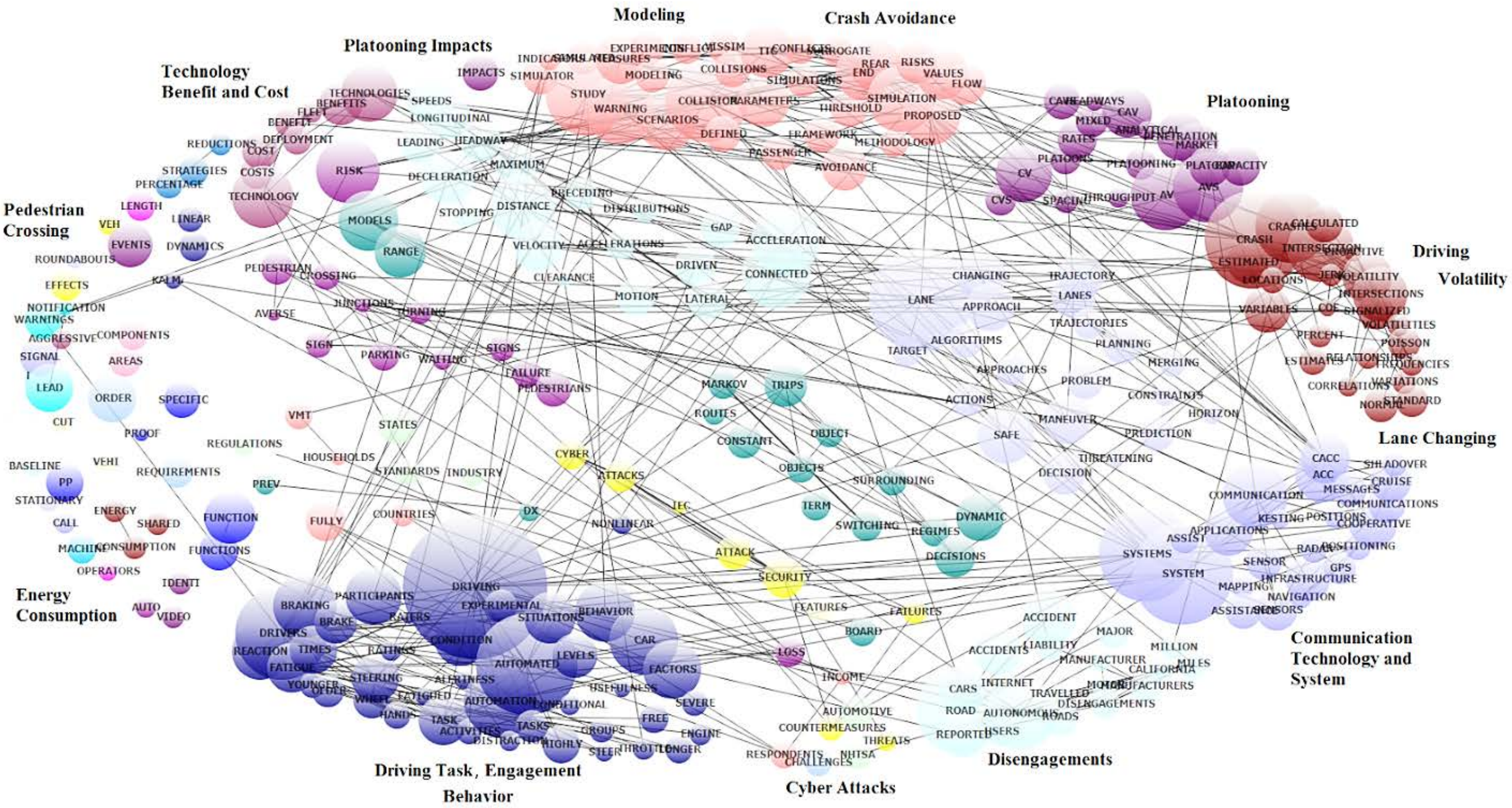

Figure 12: Concept Map of Words Across Key Topics

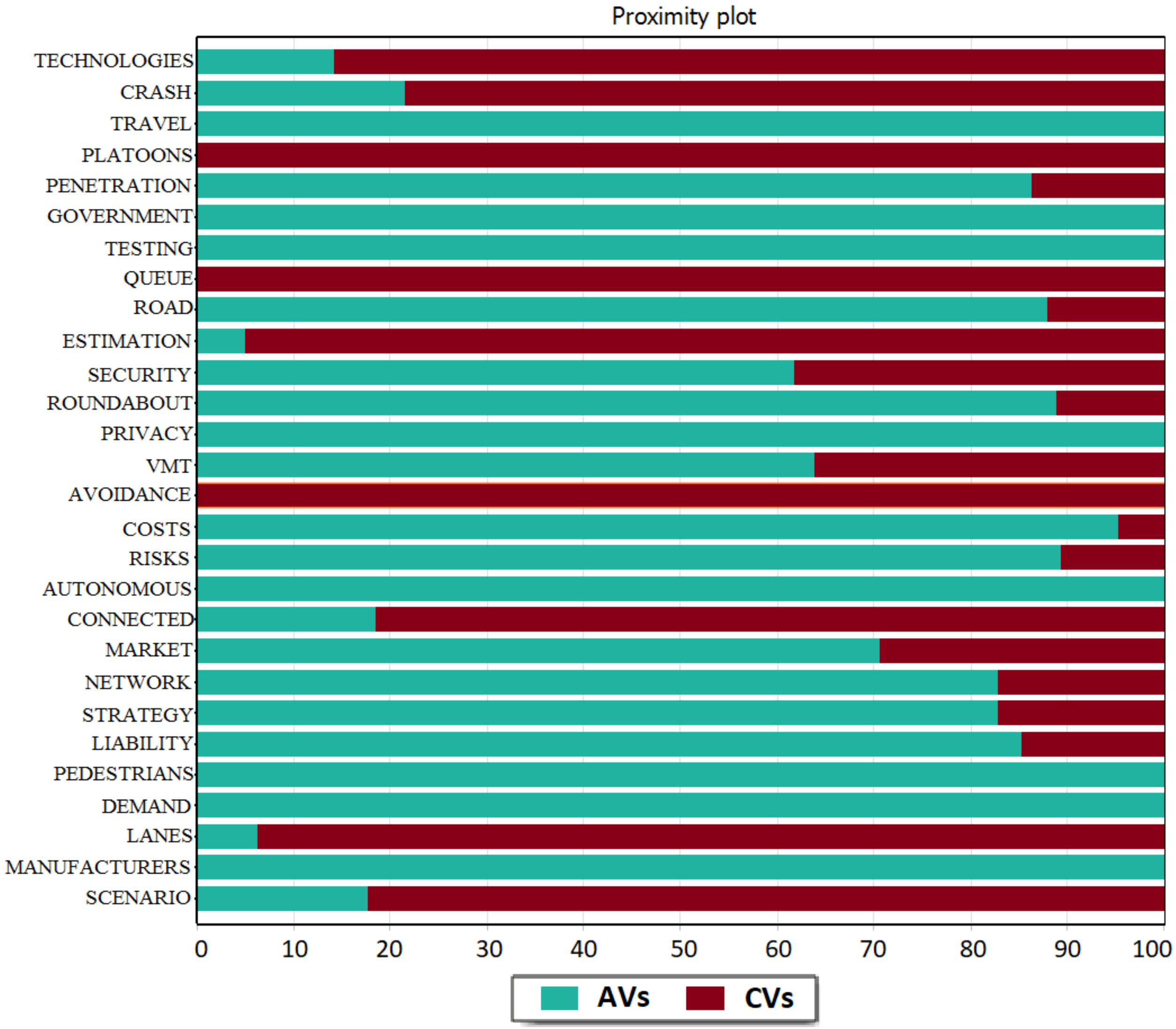

**Figure 13: Proximity Plot of AVs and CVs**

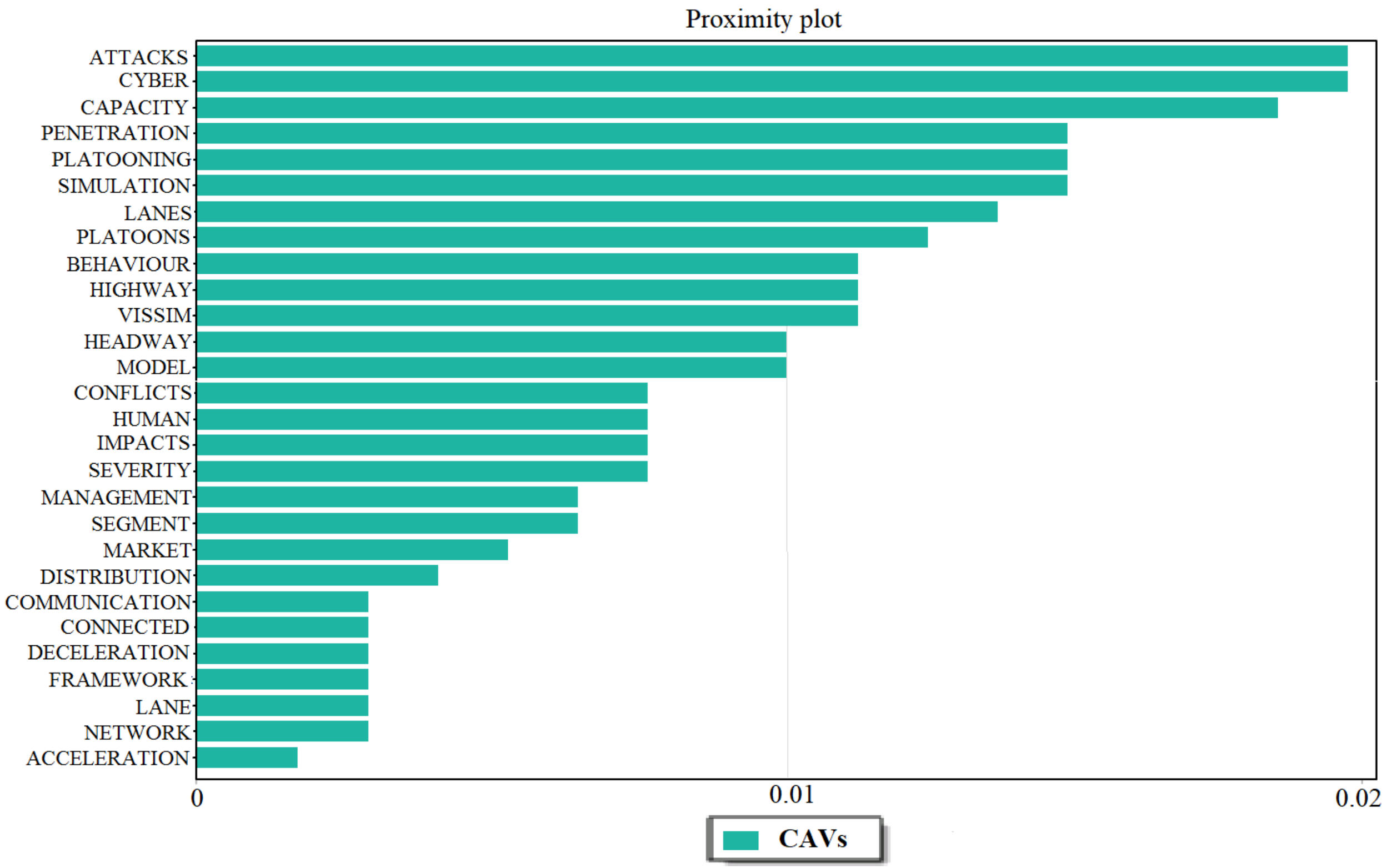

**Figure 14: Proximity Plot of CAVs**

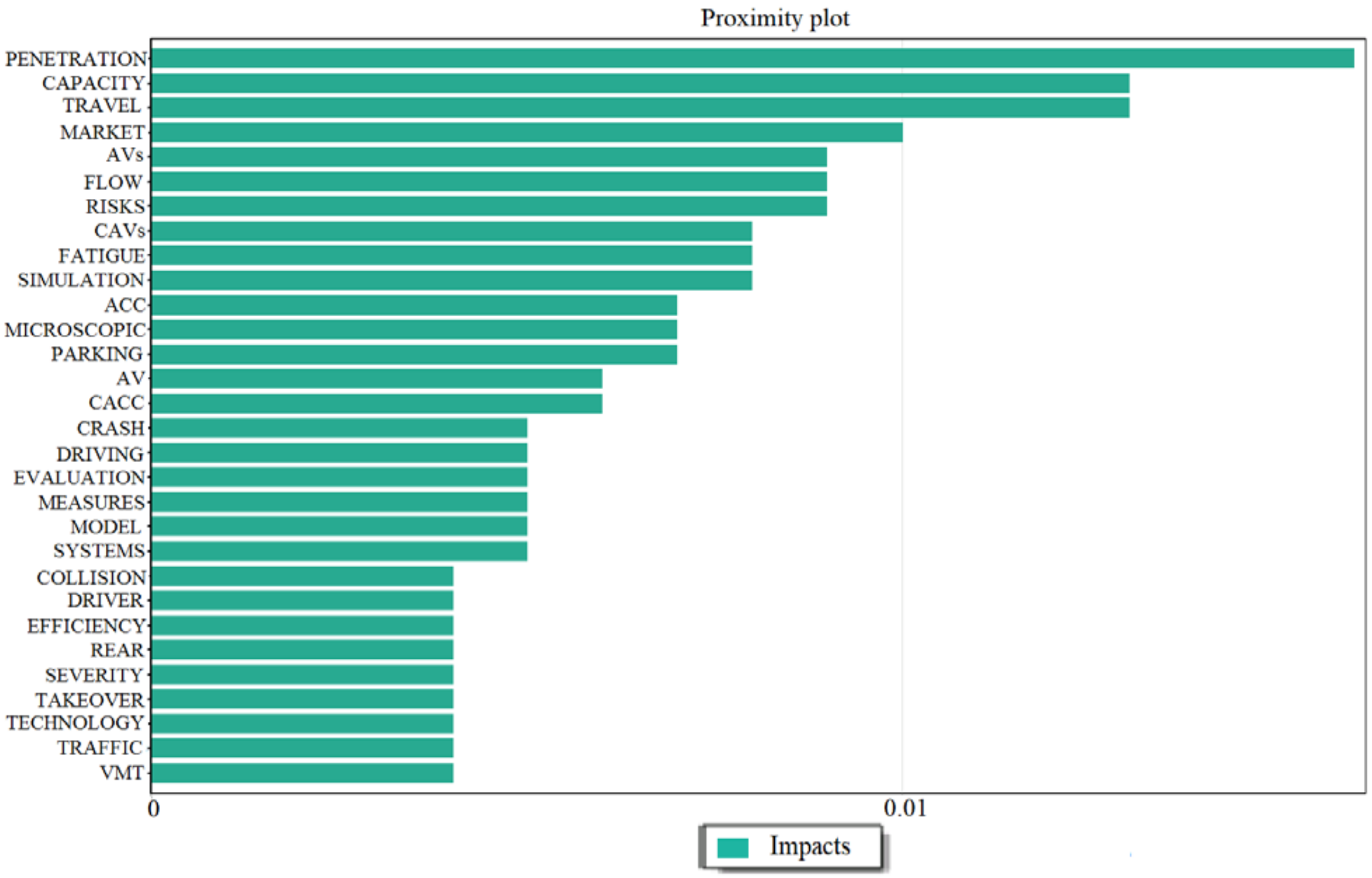

**Figure 15 Proximity Plot of Impacts**

## 6. Link Analysis and CAV Big Data Logical Architecture

Based on the outputs from previous tasks, the research team implemented predictive statistical pattern recognition techniques (such as fuzzy logic and fuzzy c-mean algorithms) to analyze connections contained in both STC's and other investigations focused on Big Data themes and concepts. Force-based graphs and 3D descriptive network graphs have been generated. The results obtained from multivariate and correspondence analysis are incorporated into link analysis. Based on the work done on Big Data by STC and other researchers, a global functional classification tree is developed, conceptualizing architecture by providing a basis for examining the concurrence of subcategories in Intelligent Transportation Systems and Connected and Autonomous Vehicle Big Data. This proves valuable in the development of a foundation for further exploring new approaches to tracking and analyzing CAV Big Data opportunities.

Based on the outputs from previous tasks, link analysis was performed to analyze interactivity and connection between Big Data themes and concepts. Figure 16 illustrates the Force-based graphs for some of the hot topics in studies conducted on Big Data for safety. These topics include: "Communication System", "Driver Behavior", "Collision Avoidance", "Lane Changing", "Market Penetration Impacts", and "Driving Volatility". In the figure, each keyword is presented by a blue node, and the interaction between them is presented by an undirected line (also called edge) in which line thickness illustrates the strength of the relationships between nodes. Red nodes represent a convergence of multiple keywords. In addition, those keywords which co-occur more often are plotted close together and vice versa. Though not quite evident in the image, this helps in the development of a foundation for further exploring new approaches to tracking and analyzing CAV Big Data opportunities.

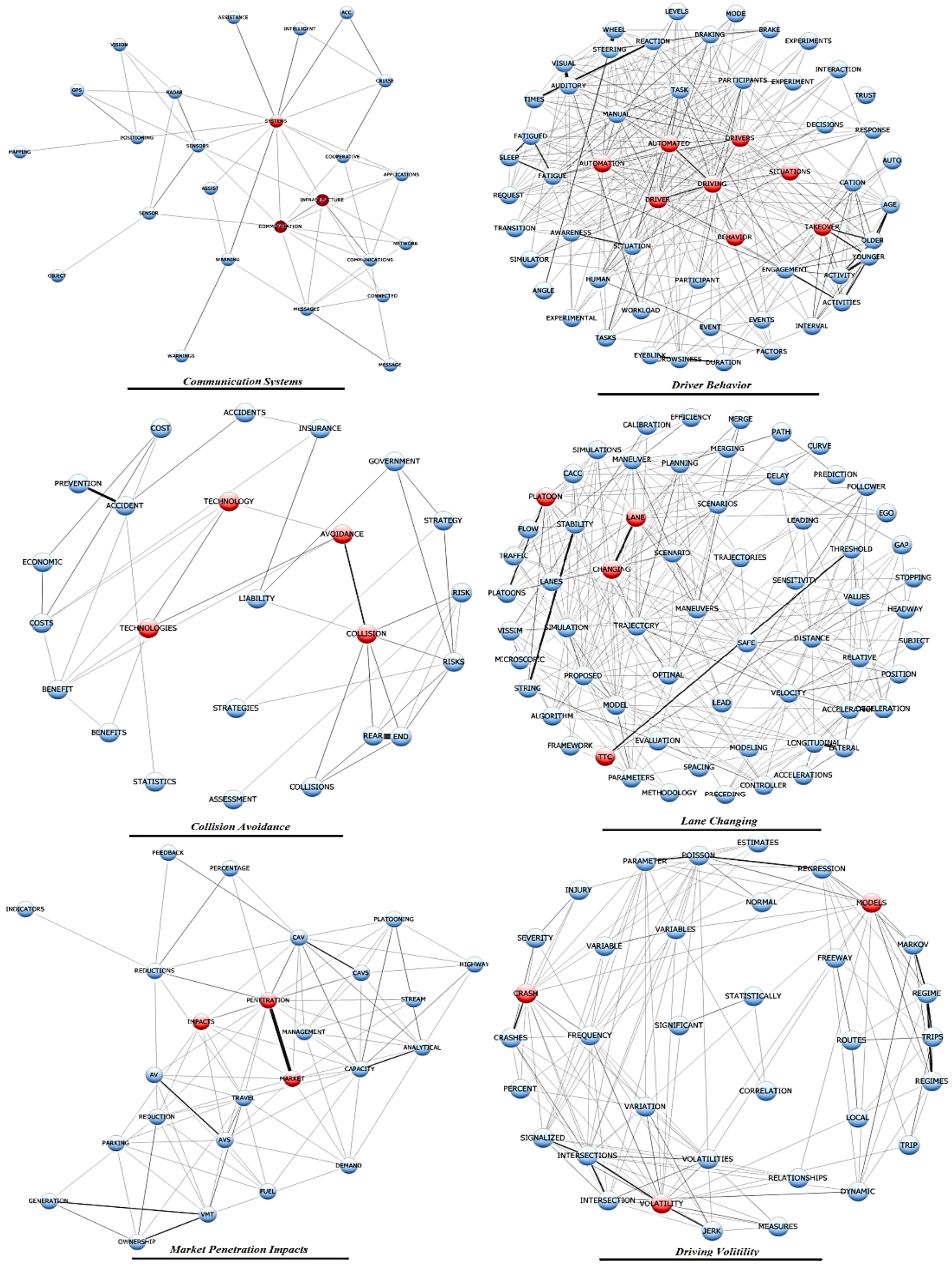

**Figure 16: Link Analysis for Selected Key Clusters**

## 7. Synthesizing and Analyzing Knowledge Generated by Southeastern Trans Center

For this task, rigorous and advanced statistical pattern algorithms, and machine learning data analytics were used to extract key patterns, high-quality information, and spot trends. Researchers developed procedures to deepen understanding by discovering new information, topics, and term relationships in a systematic manner. Specifically, in this section, the research team created a comprehensive database of Big Data related research conducted under STC's Big Data MRI. The database was created based on 14 documents; they include the reports and papers related to the STC's Big Data project (*5; 6; 117-129*). Using text mining analytics, key patterns, high-quality information, and trends are extracted. Figure 17 demonstrates the word cloud plot based on the frequency statistics of the keyword list. In a word cloud, word frequencies are converted to words with different sizes, with more frequently used words appearing larger. Also, Figure 18 shows the phrase cloud for frequency statistics of the extracted phrase list at a glance. Based on the word and phrase frequency results, most of the STC's Big Data related studies focused on big data for safety monitoring, real time crash data, travel time reliability, lane changing, and driving volatility.

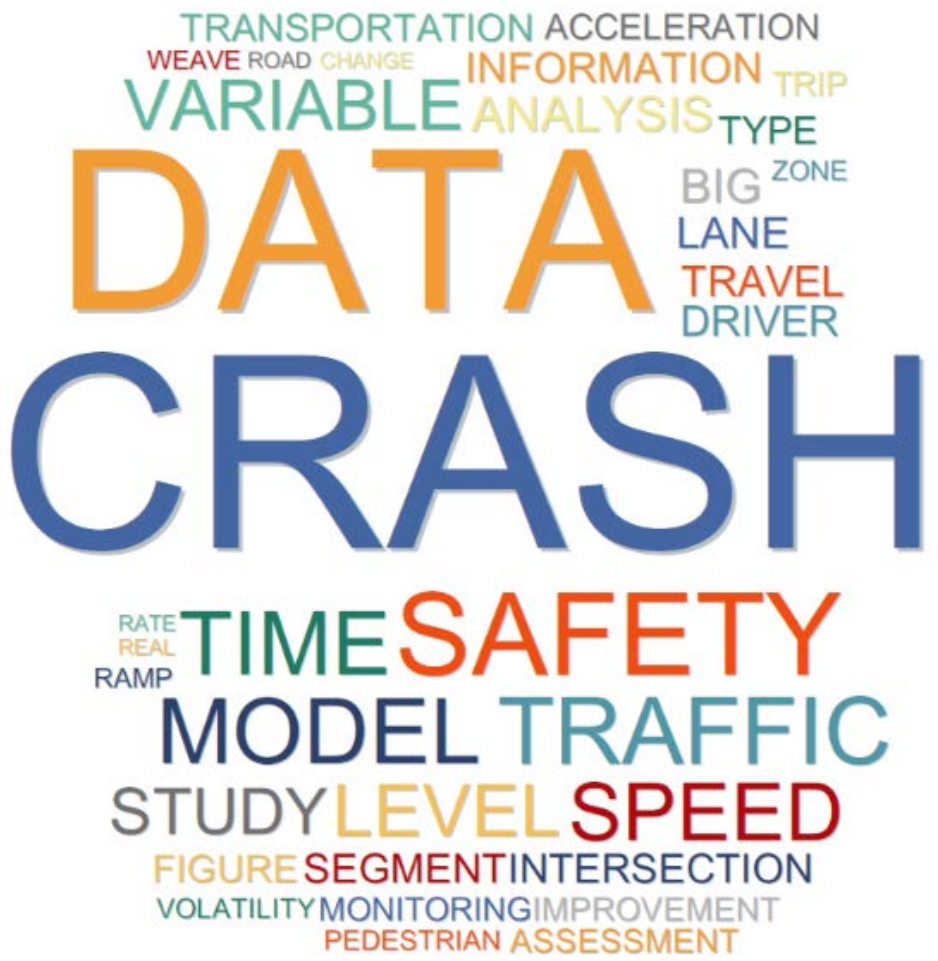

**Figure 17: Word Cloud Based on STC Studies**

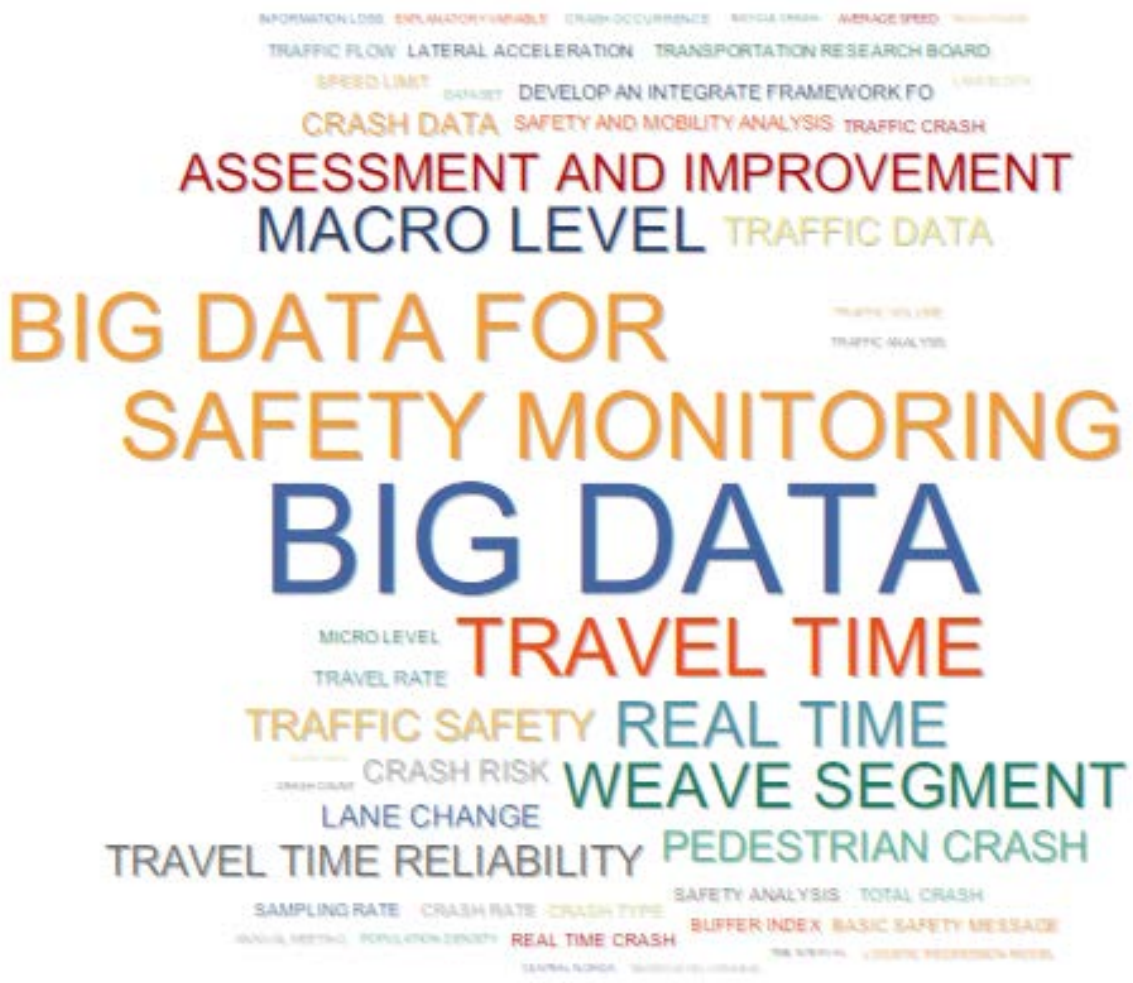

**Figure 18: Phrase Cloud Based on STC Studies**

Table 5 presents the results of topic extraction using Factor Analysis. Generally, higher eigenvalues and the percentage of coherence indicate higher variability explained by the topic meaning that the topic is more important with more coverage. Having the highest eigenvalue of 5.27, the topic of “Driving Behavior” appears in almost 91 percent of the studies. The second highest eigenvalue of 2.75 belongs to “Crash Impact on Travel Time Reliability and Real-time Crash Prediction” which appeared in almost 91 percent of the studies. Other topics of interest in descending eigenvalue order include: “Analyzing Lane Changing’ with a value of 2.26; “Macro-Level Data and Safety Analysis” with a value of 2.09, and “Big Data and Safety Monitoring” with a value of 1.76. Figure 19 provides a concept map of keywords across topics and illustrates a holistic view of the keywords and topic relationships.

**Table 5: Results of Topics Extraction through Factor Analysis (STC Studies)**

| Topic | Keywords | Coherence | Eigen value | Freq | % Cases |
|---|---|---|---|---|---|
| Driving Behavior | Decisions; Driving; Volatility; Speed; Information; Connected; Data; Big Data; Driving Decisions; Connected Vehicles; Driving Volatility; Driving Behavior; | 0.565 | 5.27 | 321 | 90.91% |
| Big Data and Safety Monitoring | Big; Safety; Data; Connected; Information; Big Data; Safety Monitoring; Connected Vehicles; Basic Safety Messages; Big Data for Safety Monitoring; | 0.431 | 1.76 | 349 | 100 % |
| Crash Impact on Travel Time Reliability, Real-time Crash Prediction | Real; Time; Crash; Traffic; Traffic Data; Traffic Safety; Time Crash; Crash Data; Expressway Ramps; Time Safety Analysis; Travel Time Reliability; | 0.523 | 2.75 | 179 | 90.91% |
| Macro-Level Data and Safety Analysis | Macro; Level; Intersection; Variables; Level Variables; Intersection Safety; Level Data; Geographic Units; Unique Database; | 0.357 | 2.09 | 124 | 81.82% |
| Analyzing Lane Changing | Lane; Change; Extreme; Extreme Lane Change Events; Lane Change Maneuvers; Normal and Extreme Lane Change; Safety Pilot Model Deployment; | 0.355 | 2.26 | 100 | 45.45% |

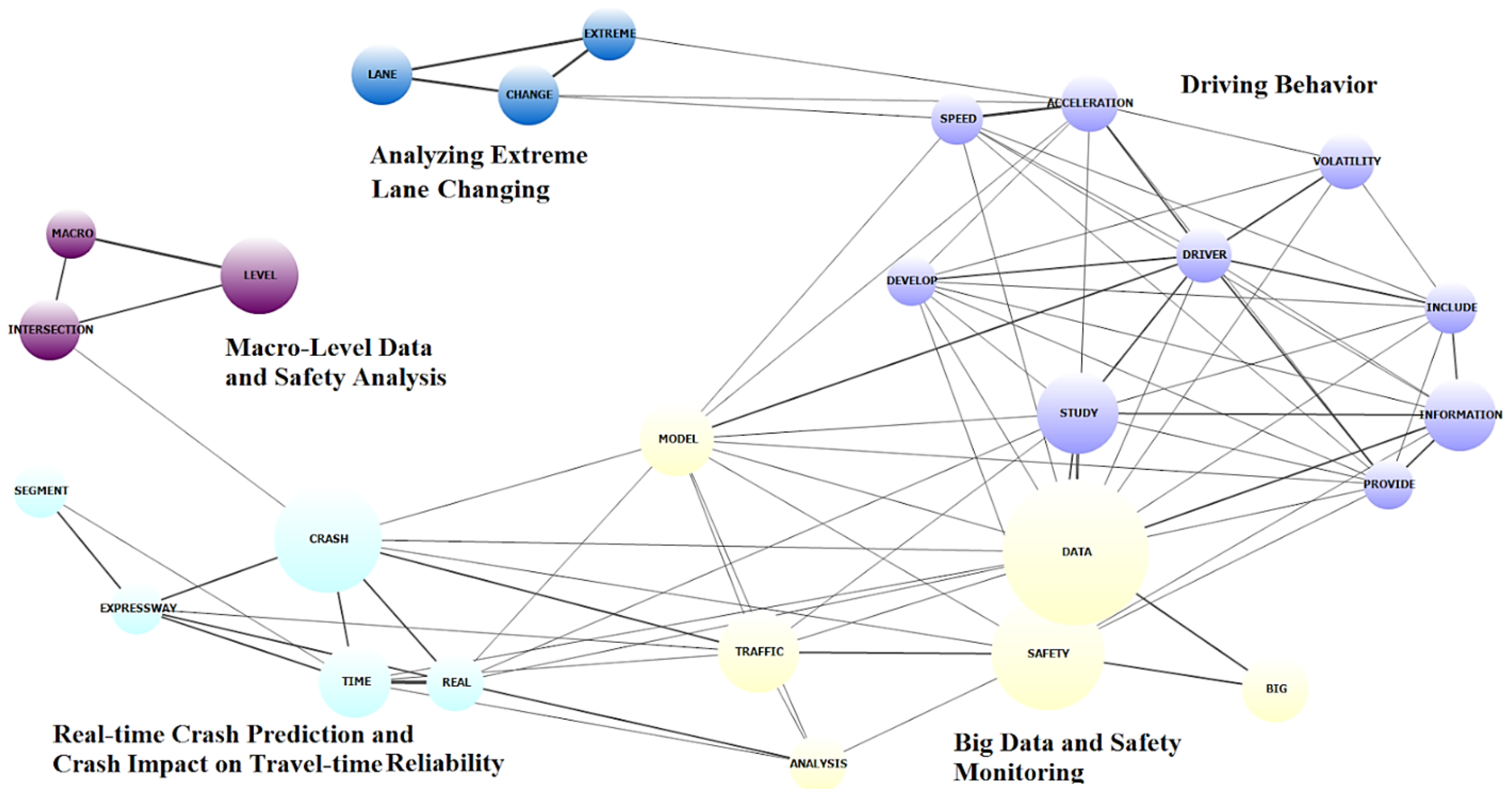

**Figure 19: Concept Map of Keywords Across Topics**

## 8. CAV Infrastructure Design Requirements

The current roadway infrastructure is designed to meet the needs of conventional vehicles driven by humans, not necessarily to meet the needs of self-driving or autonomous vehicles. Thus, the transition from conventional to self-driving, connected and autonomous vehicles might require changes to road infrastructure, such as road mapping systems, road marking, signage signalization, parking facilities, lane width, and road access points, as well as the existing requirements for new infrastructure. In this study, the research team selected 14 studies that concentrated on CAV infrastructure design to construct a text mining analysis. Table 6 summarizes the views expressed in these studies on CAV infrastructure design requirements and adjustments. Broad categories of new infrastructure requirements include concepts relating to mapping systems; road markings, signage and signalization; CAV parking needs; crossings and junctions; impacts on bridge structures; land width and road capacity; the median configuration; accessibility points, such as drop-off and pick-up points; infrastructure for fast internet connection; infrastructure for V2I communications, and regulation and legislation.

**Table 6: CAV Infrastructure Design Requirements from Selected Studies**

| Infrastructure Requirement for CAVs | | Reference |
|---|---|---|
| Mapping system | Current CAV systems strongly rely on detailed mapping of the road network. Thereby, mapping systems should include all roadworks that may alter the road layout. | (*124*) |
| | Digital infrastructure can give automated vehicles instructions from a predetermined list of road work zones. The same could be applied for law enforcement and emergency workers. | (*130*) |
| | Digital road infrastructure including highly-detailed maps and other databases about the driving environment, work zones, incidents, and traffic conditions are required to support AVs. | (*131*) |
| | One important issue in the adjustment of road infrastructure is to reflect the infrastructure in the CAVs navigation device. | (*93*) |
| Road marking, signage and signalization | CAVs performance is reliant on transparent and consistent road markings in order to navigate. Therefore, to navigate through the environment, CAVs crucially need a well-maintained road marking and signage. | (*124*) |
| | Keeping road marking in a good condition regarding the AVs requirements is necessary to provide optimal conditions for AVs on the public roads. V2I infrastructure may replace some of the functions currently performed by road signs and signals. Thus, traffic signs should be updated to enable V2I applications. | (*130*) |
| | Road marking and signage should be compatible with AVs requirements. | (*131; 132*) |
| | AVs may cause the elimination of directional signs and variable message signs. | (*133*) |

| Infrastructure Requirement for CAVs | | Reference |
|---|---|---|
| Parking for CAVs | Due to the self-park features of AVs, less car parking space is needed, which can increase parking availability. Therefore, space currently used for parking could be made available for other uses. | (*124; 134-136*) |
| | Parking facilities could be reduced, and on-street parking could be eliminated in congested areas because AVs would be able to park themselves outside of downtown or other congested areas. | (*130*) |
| Crossings and Junctions | Regarding CAVs, V2X infrastructure elements are required for pedestrian crossings, but crossings and junctions are expected to be a challenge for CAVs. | (*124*) |
| Impact on Bridge Structures | Bridge structures should be compatible with long platooning of heavy goods trucks.<br>Long platoons of trucks could change the loading on long span bridges. It would be necessary to consider this effect on the load models used in the design of structures of bridges. | (*124*) |
| Lane Width and Road Capacity | CAVs can drive in a closer distance between each other in a road. Thus, lane width could be closer to the actual vehicle width. Thereby, the same amount of traffic could be accommodated on fewer road lanes. | (*130; 133; 134*) |
| Median | Considering the CAVs features, medians could be eliminated or narrowed since a safety buffer between traffic in opposing directions is no longer required. | (*130; 131*) |
| Accessibility points (drop-off and pick-up points) | Automated vehicles could increase the need for drop-off and pick-up points due to the preference of people to be picked up and dropped off close enough to their destination. | (*130*) |
| Infrastructure for fast internet connection | Broadband deployment of fast internet connections is necessary for V2I applications. 5G technology or DSRC can be used to meet this requirement. | (*130*) (*66*) |
| Infrastructure for V2I Communication | CAVs require infrastructure to support V2I communication including both road infrastructure and user-related vehicle equipment. An example of Vehicle-to-Infrastructure communication is presented in Figure 20. | (*66; 130; 132; 137*) |
| Regulation and Legislation | Legal problems are serious obstacles for common infrastructure for conventional and autonomous vehicles. It is required to develop common rules for self-aware vehicles for all countries, making ethical decisions, and who is at fault for causing an accident. | (*93*) |

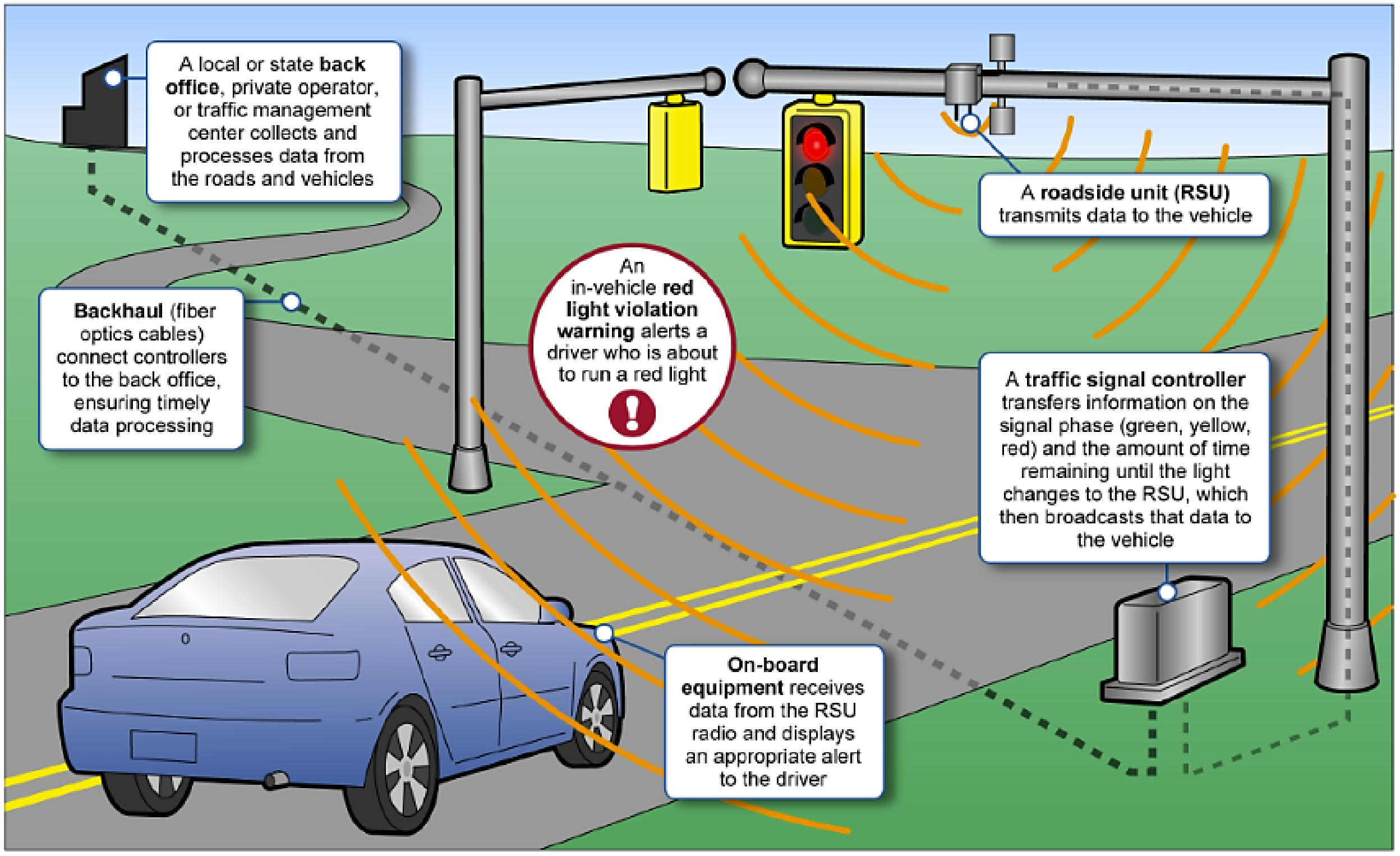

**Figure 20: Vehicle-to-Infrastructure Application Provided through Roadside Equipment (Source: GAO report; Ref 138)**

Figure 20 illustrates vehicle to infrastructure applications provided through roadside equipment provided by the Government Accounting Office in 2015 (*138*). In this figure, the vehicle has on-board equipment that received data from the surrounding infrastructure. This infrastructure includes a traffic signal controller that transfers information signal phase and timing information to a red-light warning system via a roadside unit to the vehicle, which alerts the driver to the red light. Concurrently, fiber cables connect the controllers to the traffic management center that collects and processes all data from the roads and vehicles.

The most important topics using factor analysis from the text collection regarding CAV infrastructure design are extracted. Table 7 shows the results of topic extraction. Having the highest eigenvalue value of 8.79 with a coherence of 0.376, the topic of “Security Systems” appears in almost 54 percent of the studies. The second highest eigenvalue of 3.56 belongs to “CAV Parking” with a coherence of 0.479, which appeared in all of the studies. The third highest eigenvalue (3.06) belongs to “Road Mapping, Marking, and Signs/Signals” with a coherence of 0.409, appeared in nearly 85% of the studies. The fourth highest eigenvalue of 3.00 relates to the topic of “Legislation and Regulation” with a coherence of 3.79 and was found in almost 77 percent of the literature. Other important topics in decreasing eigenvalue order include "Intersection" (2.74); "Digital Infrastructure" (2.53); "Roadside Infrastructure" (2.46), and "Database Infrastructure" (2.29).

.

**Table 7: Results of Topics Extraction through Factor Analysis – CAV Infrastructure**

| Topic | Keywords | Coherence | Eigen value | Freq | % Cases |
|---|---|---|---|---|---|
| Security Systems | Attack; Attacker; Threat; Surface; Security; System; Attack Surfaces; Autonomous Automated Vehicles; Information; | 0.376 | 8.75 | 119 | 53.85% |
| Digital Infrastructure | Radar; Camera; Navigation; GPS; Accuracy; Sensor; Map; | 0.423 | 2.53 | 127 | 84.62% |
| Roadside Infrastructure | Roadside; Unit; Fusion; Transmit; Sensor; Receive; Infrastructure; | 0.392 | 2.46 | 74 | 84.62% |
| Database Infrastructure | Waze; Incident; Live; Data; Roadwork; Database | 0.360 | 2.29 | 86 | 84.62% |
| Road Mapping, Marking, and Signs/Signals | Traffic; Road; Marking; Sign; Mark; Highway; CAV; Authority; Mapping; Traffic Management; Highway Authorities; Road Signs; Traffic Signs; Road Markings; Road Work Operators; AV Mapping Providers; Road Markings; Road Infrastructure; Road Signs and Markings; | 0.409 | 3.06 | 560 | 84.62% |
| Legislation and Regulation | Responsibility; Legislation; Law; Liability; Regulation; Regulatory; Policy; Regulate; | 0.379 | 3.00 | 103 | 76.92% |
| CAV Parking | Parking; Decrease; Park; Convert; Residential; Demand; Land; Facility; Car; Car Park; Parking Demand; Car Parking; Automated Parking; CAV Parking; Central Areas; | 0.479 | 3.56 | 303 | 100% |
| Intersections | Intersection; Dynamic; Multi; Merge; Lane; Dynamic Intersections; Status; Signal; Pedestrian; Transmit; Signal Status; Lane Capacity; | 0.386 | 2.74 | 195 | 92.31% |

The images in Figure 21 demonstrate the word and phrase cloud plots based on the frequency statistics of the word and phrase list regarding CAV infrastructure design. This plot illustrates the most frequent words and phrases which appear in the related studies. In a word cloud, frequencies are converted to words with different sizes, the more frequently the word appeared in the studies, the larger the word or phrase would be in the plot. For example, the word cloud highlights the words Automate, System, Road, Traffic, CAV, Infrastructure and Car, whereas, the phrase cloud highlights the phrases Car Park, Traffic Management, Highway Authority, Traffic Sign, and Road Marking.

From the results, although of vital importance, few studies have been conducted on infrastructure design requirements. The relevant studies mostly address parking facilities, road marking and signing, traffic signalization, and mapping systems. However, these topics still need to be explored more comprehensively in detail to anticipate future CAV infrastructure design requirements. Moreover, numerous crucial issues need to be studied and addressed. These issues include roadway structures (e.g. bridges, and pavement design), and geometric design.

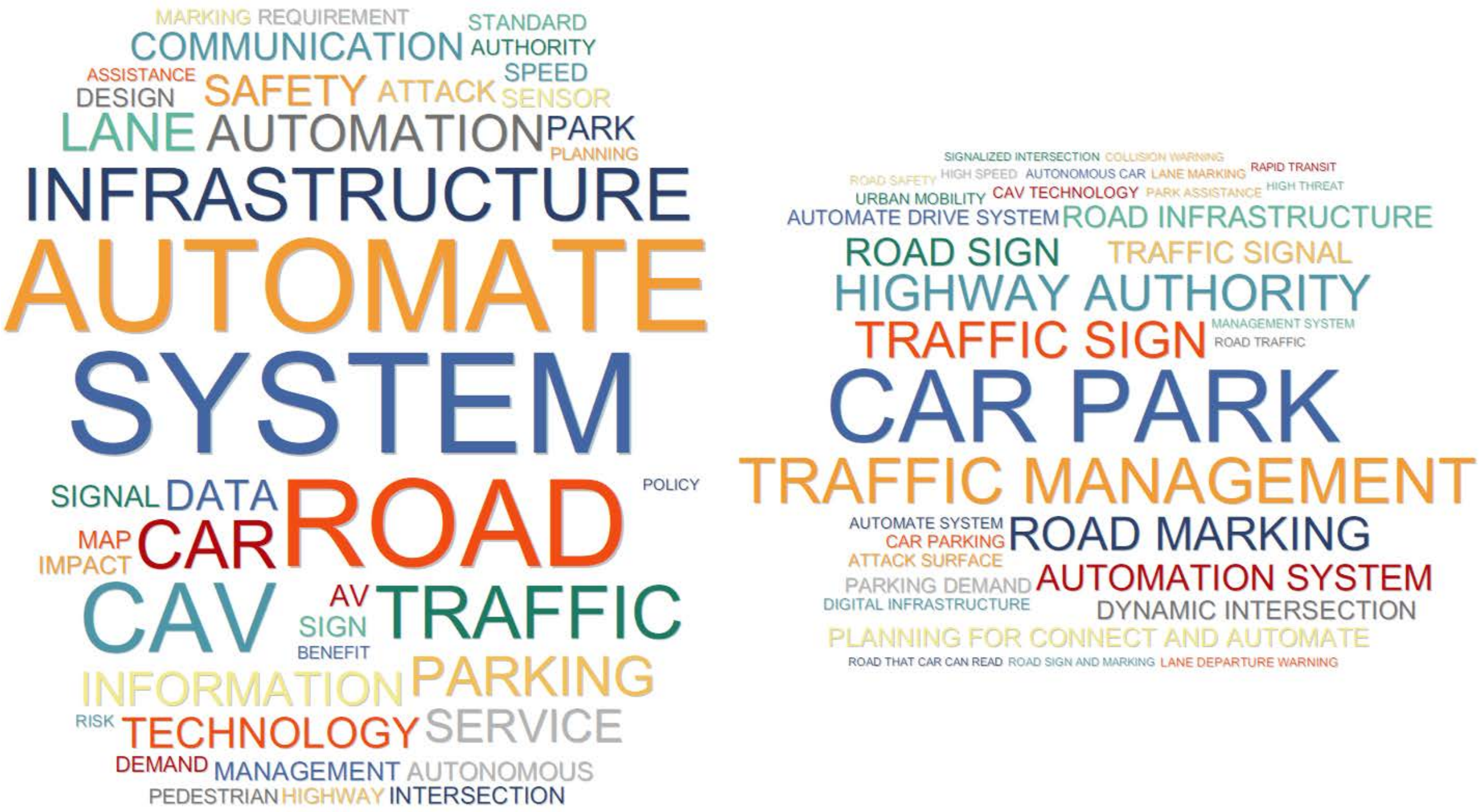

**Figure 21: Word and Phrase Clouds for CAV Infrastructure**

## 9. Conclusions

Rapid technological advancements in recent years have manifested in the emergence of connected and automated vehicles. The next frontier of transportation development is to equip motor vehicles and transportation systems with wireless communication technologies in a bid to establish cooperative, well-informed, and proactive transportation systems. In this context, it is important to consider transformational advancements in transportation-related-data-collection techniques that acquire increasing amounts of information generated by electronic sensors in combined ecosystems of CAVs. While generated large-scale empirical data has significant potential to facilitate a deeper understanding of transportation-related problems, integrating and processing large datasets in a meaningful manner is still an open challenge. To this end, the Southeastern Transportation Center's "Big Data" major research initiative, which focuses on developing innovative frameworks for generating useful knowledge from fragmented, disorganized, and difficult-to-analyze "Big Data," is of critical importance.

The study utilized different search engines to generate more than 100 documents including CAV Big Data and safety along with infrastructure design, and STC Big Data initiatives. Rigorous and advanced machine learning text analytics tools are utilized to extract key concepts, spot patterns, and trends in STC sponsored Big Data initiatives and related studies. The study synthesizes and derives high-quality information from innovative research activities undertaken by various research entities through Big Data initiatives.

To answer the fundamental question of how CAVs will improve road users' safety, this report focuses on data generated from connected and automated vehicles. By using advanced data mining and thematic text analytics tools, the goal is to systematically synthesize studies related to Big Data for safety monitoring and improvement. Within this domain, the report compares Big Data initiatives nationally and internationally and provides insights regarding the evolution of Big Data science applications related to CAVs.

Big data are being harnessed for driver assistance (alerts and warnings) as well as to enable higher levels of connectivity and automation where vehicles can take over control of speed and steering. The specific technologies that are being discussed in the literature include collision avoidance and vehicle control. Based on the text mining results, promising technologies include "Collision Avoidance and Warning", "Collision Control", "Communication Systems and Data", "Autonomous Lane Changing", and "Driving Task, Engagement/ Disengagement". Topics that are most frequently discussed in the literature and cover the context include "Market Penetration Impacts", "Driving Volatility", "Driver Behavior", "Big Data", and more broadly "Intelligent Transportation Systems". These results help provide a conceptual foundation for developing new approaches for guiding and tracking the safety implications of Big Data and related innovations for CAVs. Although CAVs have substantial potential to improve traffic safety, the results reveal new challenges that CAV technology should address, such as cyber-attacks, privacy violation, invalidated assumptions, uncertain interaction behavior (e.g., between automated vehicles and pedestrians or bicyclists), technology failure, driver in attention (that can increase crash risk due to disengagement), high technology cost, and regulation inconsistency across states.

The study points out that driving errors can be potentially be avoided with emerging CAV technologies, resulting in safer transportation system operation. However, the safety of vehicle automation systems is still being researched and tested. Higher level automation is in early stages of development. Some drivers are uncomfortable with relinquishing control of their vehicles, especially in complex urban environments. At vehicular speeds, occupants' lives can depend on the performance of sensors and data that is being transmitted. Hence, creating, processing, and analyzing such big data generated by vehicles in real-time is critical for the success and safe operation of CAVs. Data stream analytics and identifying unusual events in data streams is a new dimension of big data that was identified as critically important going forward.

To reap a wide range of CAV benefits, infrastructure design should become more CAV-friendly. Although of vital importance, few studies have been conducted on infrastructure design requirements. The relevant studies mostly address parking facilities, road marking and signing, traffic signalization, and mapping systems. However, these topics still need to be explored more comprehensively in detail to guide future CAV infrastructure design requirements. Moreover, numerous crucial issues need to be studied and addressed, such as roadway structure (e.g. bridges, and pavement design), and geometric design.

A wide range of stakeholders from government agencies, private sector companies, and academia are involved in CAV research, development, and deployment. They are also concerned about ensuring that CAVs are safe. Clearly the sensors, machine intelligence and algorithms that create and process big data, especially for higher levels of vehicle automation, should be certified. In this regard, more research and development supported by the stakeholders will be critical to share safety practices and harness the true potential of CAV data. The topics needing more research are identified in this report, especially, analytics for archived and real-time CAV data, understanding safety data that can eliminate transportation injuries and deaths, and changes to infrastructure geometry needed for successfully facilitating CAV operation.

Limitations regarding the results of this study are that they reflect the inputs; this means the results are affected by the selected literature and searched keywords. Nevertheless, new knowledge generated under the STC Big Data MRI together with other initiatives can ultimately help in developing a foundation for guiding future research initiatives and developing new procedures for tracking and analyzing CAV Big Data and related innovations.